\documentclass[aps,prd,twocolumn,superscriptaddress,showpacs,letterpaper]{revtex4}

\usepackage{verbatim}
\usepackage[usenames,dvipsnames]{color}
\usepackage[pdftex]{graphicx}
\usepackage[english]{babel}
\usepackage{amsmath,amssymb}
\usepackage[colorlinks=true, citecolor=blue, linkcolor=WildStrawberry]{hyperref}

\usepackage{multirow}
\usepackage{mathptmx}

\begin{document}

\newcommand{\E}{\mathrm{E}}
\newcommand{\Var}{\mathrm{Var}}
\newcommand{\bra}[1]{\langle #1|}
\newcommand{\ket}[1]{|#1\rangle}
\newcommand{\braket}[2]{\langle #1|#2 \rangle}
\newcommand{\mean}[2]{\langle #1 #2 \rangle}
\newcommand{\be}{\begin{equation}}
\newcommand{\ee}{\end{equation}}
\newcommand{\ba}{\begin{eqnarray}}
\newcommand{\ea}{\end{eqnarray}}
\newcommand{\SD}[1]{{\color{magenta}#1}}
\newcommand{\rem}[1]{{\sout{#1}}}
\newcommand{\alert}[1]{\textbf{\color{red} \uwave{#1}}}
\newcommand{\Y}[1]{\textcolor{yellow}{#1}}
\newcommand{\R}[1]{\textcolor{black}{#1}}
\newcommand{\B}[1]{\textcolor{black}{#1}}
\newcommand{\C}[1]{\textcolor{cyan}{#1}}
\newcommand{\db}{\color{darkblue}}
\newcommand{\intinfty}{\int_{-\infty}^{\infty}\!}
\newcommand{\Tr}{\mathop{\rm Tr}\nolimits}
\newcommand{\const}{\mathop{\rm const}\nolimits}
\makeatletter
\newcommand{\rmnum}[1]{\romannumeral #1}
\newcommand{\Rmnum}[1]{\expandafter\@slowromancap\romannumeral #1@}

\makeatother

\title{Brownian Thermal Noise in Multilayer Coated Mirrors}
\author{Ting Hong}
\affiliation{Theoretical Astrophysics 350-17, California Institute of Technology, Pasadena, CA 91125, USA}
\author{Huan Yang}
\affiliation{Theoretical Astrophysics 350-17, California Institute of Technology, Pasadena, CA 91125, USA}
\author{Eric K. Gustafson}
\affiliation{LIGO Laboratory 100-36, California Institute of Technology, Pasadena, CA 91125, USA}
\author{Rana X. Adhikari}
\affiliation{LIGO Laboratory 100-36, California Institute of Technology, Pasadena, CA 91125, USA}
\author{Yanbei Chen}
\affiliation{Theoretical Astrophysics 350-17, California Institute of Technology, Pasadena, CA 91125, USA}
\date{\today}

\begin{abstract}
We analyze the Brownian thermal noise of a multi-layer dielectric coating, 
used in high-precision optical measurements including interferometric gravitational-wave 
detectors.  We assume the coating material to be isotropic, and therefore study thermal 
noises arising from shear and bulk losses of the coating materials.  We show that coating noise 
arises not only from layer thickness fluctuations, but also from fluctuations of the interface 
between the coating and substrate, driven by internal fluctuating stresses of the coating. In
addition, the non-zero photoeleastic coefficients of the thin films modifies the influence of
the thermal noise on the laser field.
The thickness fluctuations of different layers are statistically independent, however, 
there exists a finite coherence between layers and the substrate-coating interface.  Taking into account 
uncertainties in material parameters, we show that significant uncertainties still exist in
estimating coating Brownian noise.
\end{abstract}

\maketitle


\section{Introduction}
Brownian thermal noise in the dielectric coatings of mirrors limits some high precision 
experiments which use optical metrology. This thermal noise is currently a limit for
fixed spacer Fabry-Perots used in optical clock experiments~\cite{JunYe} and is
estimated to be the dominant noise source in the most sensitive band of modern gravitational wave
detectors (e.g., advanced LIGO, GEO, advanced VIRGO and LCGT)~\cite{LIGO:RPP,aLIGO,GEO,AdV, LCGT}. Recent 
work has indicated the possibility of  suppressing the various kinds of thermal noise 
by redesigning the shape of the substrate and the structure of the multi-layer 
coating~\cite{Kimble, Evans}. 
In this paper, we seek a comprehensive understanding of 
coating Brownian noise, by identifying individual sources
of fluctuations, calculating their cross spectra using the 
fluctuation dissipation theorem~\cite{Callen:FDT, Gonzalez:JASA, Levin:Direct}, 
and finally evaluating how each of the sources and their correlations
add up to the total noise.

As a starting point, we will assume each coating layer to be isotropic, and hence 
completely characterized by its complex bulk modulus $K$ and shear modulus 
$\mu$---each with small imaginary parts related to energy loss in bulk and shear 
motions.  The complex arguments of these moduli are often referred to as {\it loss angles}. 
While values of $K$ and $\mu$ are generally known, loss angles vary significantly 
according to the details of how coating materials are applied onto the substrate and their composition.  Since the 
loss angles are small, we will still use $K$ and $\mu$ to denote the real parts of the bulk 
and shear moduli, and write the complex bulk and shear moduli, $\tilde K$ and $\tilde \mu$ as
\begin{equation}
\tilde K = K (1+i\phi_B)\,,\quad \tilde\mu = \mu (1+i\phi_S)\,.
\end{equation}
Here we have used subscripts $B$ and $S$ to denote bulk and shear, because these will 
be symbols for bulk strain and shear strain. Note that this definition differs from previous 
literature and measurements, which used $\phi_\parallel$ and $\phi_\perp$  to denote losses 
induced by elastic deformations parallel and orthogonal to the coating-substrate 
interface~\cite{harry}. We argue in Appendix~\ref{App:lossangle} that $\phi_\parallel$ and $\phi_\perp$ cannot be consistently used as the loss angles of a material.

\begin{figure}
   \includegraphics[width=2.75in]{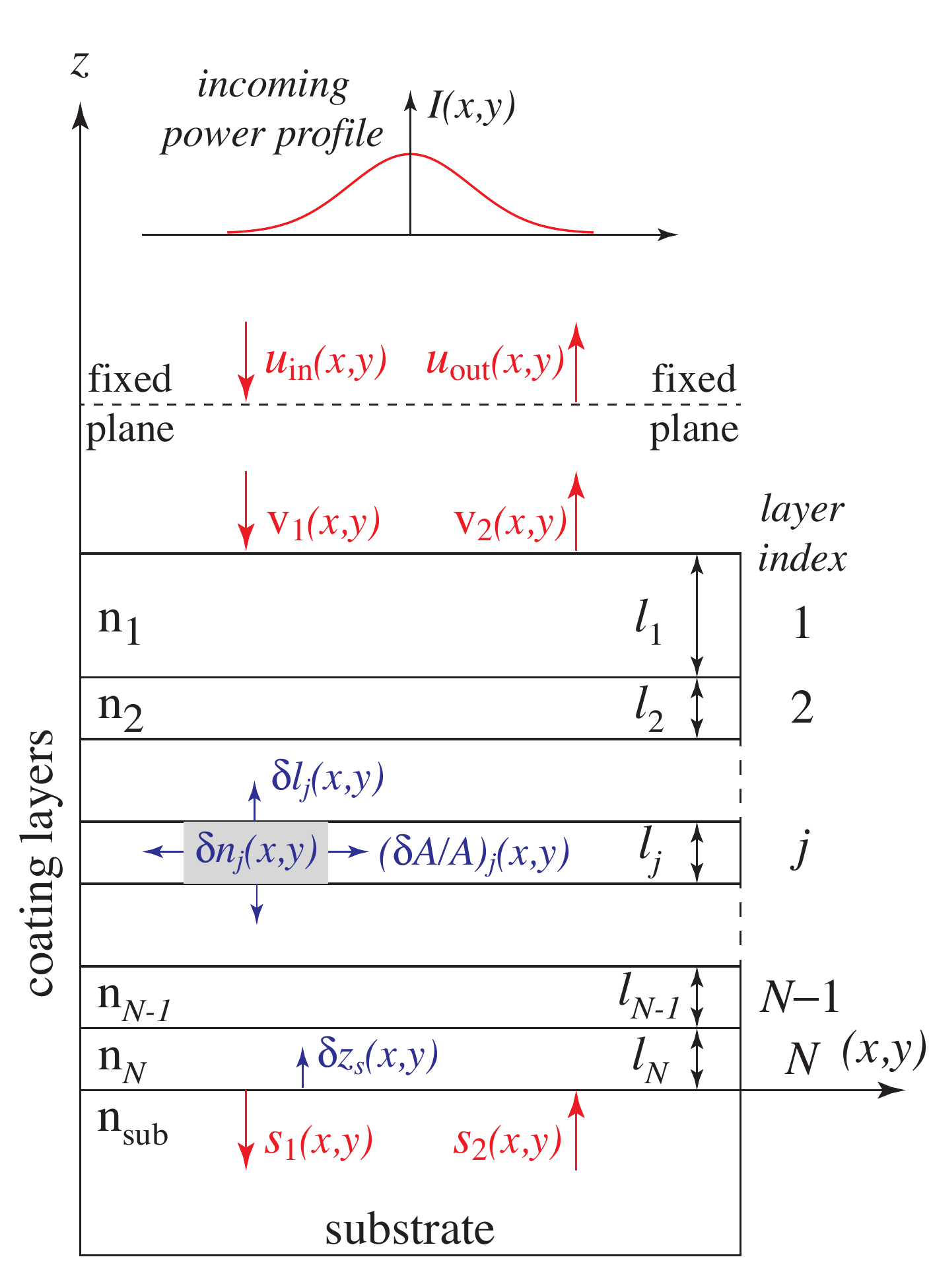}
    \caption{
     Schematic plot of a mirror coated with multiple dielectric layers. 
     Shown here are the various fluctuations that contribute to coating noise, i.e., fluctuations 
     in the amplitude and phase of the returning light caused by fluctuations in the geometry (e.g. 
     layer thickness $\delta l_j$, layer area stretch $(\delta A/A)_j$ and interface height $z_s$) of the 
     coating-substrate configuration and in refractive indices [$\delta n_j(x,y,z)$] of the layers.} 
  \label{fig:coating_config}
\end{figure}

Brownian thermal fluctuations of a multilayer coating can be divided as follows: 
(i) thickness fluctuation of the coating layers, (ii) fluctuation of the coating-substrate interface, and (iii) 
refractive index fluctuations of the coating layers associated with longitudinal (thickness) and 
transverse (area) elastic deformations---as illustrated in Figure~\ref{fig:coating_config}.  Using 
what is sometimes referred to as the Levin's direct approach~\cite{Levin:Direct} (based on the fluctuation dissipation theorem), one can lump all three 
contributions into one quantity, and calculate its noise spectral density by calculating the 
mechanical dissipation rate when a distribution of mechanical forces are applied at various locations on 
the coated mirror [as has been done by Vyatchanin et al.~\cite{sergey}].  However, in order to 
obtain insights into coating noise that have proven useful we have 
chosen to calculate the cross spectral densities for each of (i), (ii), and (iii), and provide intuitive interpretations of each.  
We will show, that (i) and (ii) above are driven by both bulk 
and shear fluctuations in the coating, in such a way that thickness fluctuations of the $j$-th 
layer $\delta l_j$, or in transverse locations separated by more than coating thickness, are mutually 
statistically independent, yet each $\delta l_j$ is correlated with the fluctuation of the 
coating-substrate interface $z_s$---because $z_s$ is driven by the {\it sum} of thermal stresses 
in the coating layers.  We will also show that when coating thickness is much less than the beam spot
size, the only significant contribution to (iii) arises from longitudinal (thickness) fluctuations.

This paper is organized as follows.  In Sec.~\ref{sec:components}, we express the amplitude and phase of the out-going field
in terms of fluctuations in the coating structure, thereby identifying the various components of coating thermal noise.  
In Sec.~\ref{sec:nopenetration}, we introduce the loss angles of isotropic coating materials, and use the Fluctuation-Dissipation Theorem to calculate the cross spectral densities 
of the coating thermal noise ignoring light penetration into the 
multi-layer coating. In Sec.~\ref{sec:spectra}, we discuss in detail the cross spectra of all the components of the coating structure fluctuation, thereby obtaining the full formula for coating thermal noise, taking light penetration into account.  In Sec.~\ref{sec:penetration}, we discuss the effect of light penetration on coating thermal noise, using typical optical coating structures.  
In Sec.~\ref{sec:dependence}, we discuss the dependence of thermal noise on the material parameters, and optimize the coating structure in order to lower the thermal noise. In Sec.~\ref{sec:measurement}, we discuss how only one combination of the two loss angles have been measured in past experiments, and how other different combinations  can be measured using a new experimental geometry. Finally, we summarize our main conclusions in Sec.~\ref{sec:conclusions}.

%
%

\section{Components of the coating thermal noise}
\label{sec:components}
In this section, we express the coating thermal noise in terms of the elastic deformations of the coated substrate. 

\subsection{Complex Reflectivity}
As illustrated in Figure~\ref{fig:coating_config}, we consider a laser field normally incident 
(along the $-z$ direction) onto the mirror, with complex amplitude profile $u_{\rm in}(x,y)$ at a 
fixed reference plane (dashed line in the figure) and intensity profile $I(x,y)=|u_{\rm in}(x,y)|^2$. 
Henceforth in the paper, we shall use arrows (e.g., $\vec x$) to denote the 2-dimensional vector 
$(x,y)$ in the {\it transverse plane}, and boldface letters (e.g., $\mathbf{x}$) to denote 3-dimensional vectors. 

Because coating thickness is much less than the beam spot size, the reflected field (traveling along the $+z$ direction) at transverse location $\vec x$ has an amplitude given by
\begin{equation}
u_{\rm out} (\vec x) = \rho_{\rm tot} (\vec x)  u_{\rm in}(\vec x)\,,
\end{equation}
%
%
which only depends on the complex reflectivity $\rho_{\rm tot}(\vec x)$ and the complex amplitude of the incident field $u_{\rm in}(\vec x)$, at the same location $\vec x$ --- assuming no incident light from the substrate (i.e., $s_2=0$).  Here $\rho_{\rm tot}(\vec x)$ can be separated into three factors, as
\begin{equation}
\label{r1}
\rho_{\rm tot}(\vec x) = \frac{u_{\rm out}(\vec x)}{u_{\rm in}(\vec x)} =
\left[\frac{u_{\rm out}(\vec x)}{v_2(\vec x)}\right]
\left[\frac{v_1(\vec x)}{u_{\rm in}(\vec x)}\right]
\left[\frac{v_2(\vec x)}{v_1(\vec x)}\right]
\end{equation} 
%
%
in which $v_1(\vec x)$ is the incident complex amplitude at the coating-air interface, while $v_2(\vec x)$ 
is the reflected complex amplitude at that interface.

\R{The first two phase factors on the right-hand side of Eq.~\eqref{r1} are gained by the light
when traveling across the gap between the fixed reference plane (see Fig.~\ref{fig:coating_config}) and the coating-air interface; we therefore obtain, up to a constant phase factor, }
\R{\begin{equation}
\label{r2}
\left[\frac{u_{\rm out}(\vec x)}{v_2(\vec x)}\right]
\left[\frac{v_1(\vec x)}{u_{\rm in}(\vec x)}\right]=e^{-2 i k_0 \left[\delta z_s(\vec x)+ \sum_{j=1}^N \delta l_j(\vec x)\right]}
\end{equation}}
where $k_0 = \omega_0/c$ is the wave number of the laser ($\omega_0$ its angular frequency) in 
vacuum, $z_s(\vec x)$ is the vertical displacement of the coating-substrate interface (from its zero point), 
and $l_j (\vec x)$ is the thickness of the $j$-th coating layer --- both evaluated at a transverse location 
$\vec x$.  

The remaining complex reflectivity $v_2(\vec x)/v_1(\vec x)$ can be determined as a function 
of the phase shift experienced by the field in each layer, as well as the reflectivity of each interface, 
as described in detail in Sec.~\ref{sec:penetration}.  We can write:
\begin{equation}
\label{r3}
v_2/v_1 = \rho[\phi_1(\vec x),\ldots,\phi_N(\vec x);r_{01}(\vec x),\ldots,r_{Ns}(\vec x)]
\end{equation}
Here $\rho$ is the complex reflectivity of a multi-layer coating, measured at the coating-air interface, 
which in turn depends on the optical thickness $\phi_j(\vec x)$ of each layer ($j=1,\ldots,N$) and the 
reflectivity $r_{p,p+1}(\vec x) \equiv r_p(\vec x)$ of each interface, ($p=0,\ldots,N$, with $p=N+1$ 
representing the substrate, and $p=0$ the vacuum outside the coating).  
Assembling the above equations \eqref{r1}--\eqref{r3}, we obtain:
\begin{equation}
\label{rhotot}
\rho_{\rm tot}(\vec x) = e^{-2 i k_0 \left[\delta z_s(\vec x)+ \sum_{j=1}^N \delta l_j(\vec x)\right]}\rho[\{\phi_j(\vec x)\};\{r_{p}(\vec x)\}]
\end{equation}
Brownian thermal forces lead to fluctuations in both the real and imaginary parts of this complex reflectivity.
Fluctuations in the argument of the complex reflectivity phase modulates the out-going light and directly produces sensing noise.
Fluctuations in the magnitude, on the other hand, amplitude modulate the out-going light, and produces a ponderomotive force noise. 

%
%

\subsection{Thermal Phase and Amplitude Noise}
\label{subsec:pham}

Brownian thermal fluctuations in coating geometry and refractive index modify the total reflectivity $\rho_{\rm tot}(\vec x)$  defined in Eq.~\eqref{rhotot}. \R{The real and imaginary parts of 
\begin{equation}
\delta\log\rho_{\rm tot}(\vec x) =\frac{\delta\rho_{\rm tot}(\vec x)}{\rho_{\rm tot}(\vec x)}
\end{equation}
encode the amplitude/intensity and phase fluctuations of the reflected light at position $\vec x$ on the mirror surface. In particular, intensity fluctuation of the reflected light is given by
\begin{equation}
\frac{\delta I(\vec x)}{I(\vec x)} = 2\frac{\delta |\rho_{\rm tot} (\vec x)|}{|\rho_{\rm tot} (\vec x)|}  = 2 \mathrm{Re} \left[\delta \log\rho_{\rm tot} (\vec x) \right]
\end{equation} 
while phase fluctuation is given by
\begin{equation}
\label{eq:phix}
\delta\phi(\vec x) =  \delta\arg\left[\rho_{\rm tot}(\vec x)\right]  =  \mathrm{Im} \left[\delta \log\rho_{\rm tot}(\vec x)\right]
\end{equation}
In this way, if we} further write
\begin{equation}
\label{rho_re_im}
\xi(\vec x) - i  \zeta(\vec x) =-\frac{i}{2k_0}\delta \left[\log\rho_{\rm tot} \right],
\end{equation}
\R{with both $\xi$ and $\zeta$  real-valued functions of $\vec x$, with the dimensionality of displacement; they will represent phase and amplitude noise, respectively.  
In particular, from Eq.~\eqref{eq:phix}, we have  
\begin{equation}
\label{xi}
2k_0 \xi(\vec x)  = \delta \phi(\vec x)\,,
\end{equation}
and this means $\xi(\vec x)$ corresponds to the spurious displacement measured by the reflected light due to phase fluctuations caused by the coating.}

\R{The quantity $\zeta$  is connected to amplitude/intensity noise via 
\begin{eqnarray}
\label{ampth}
2 k_0\zeta(\vec x) = \mathrm{Re}\left[\delta \log\rho_{\rm tot}\right] = \frac{\delta I(\vec x)}{2I(\vec x)}\,.
\end{eqnarray}
As we shall discuss in Sec.~\ref{subsec:amp}, $\zeta$ will cause a fluctuating force on the mirror, and contribute to measurement noise, although the effect will be small for gravitational-wave detectors.}

%
%
%
%
%

Inserting the dependence of $\rho_{\rm tot}$ on $\rho$, $l_j$ and $z_s$, we obtain 
\begin{eqnarray}
\label{coth}
\xi(\vec x) -i\zeta(\vec x) &=&
-\delta z_s(\vec x) -
\sum_{l=1}^N \delta l_j(\vec x)
\nonumber \\& -&
\sum_{j=1}^N
\frac{i}{2 k_{0}}
\left[
\frac{\partial \log \rho}{\partial \phi_j}\cdot \delta\phi_{j} (\vec x)\right] \nonumber\\
&-&\sum_{p=0}^N \frac{i}{2k_0}\,  \left[\frac{\partial \log \rho}{\partial r_p}\cdot\delta r_p (\vec x)\right]\,.
\end{eqnarray}
The first two terms are due to the motion of the coating-air interface at location $\vec x$ and thickness fluctuations of 
the layers, while the last two terms are due to light penetration into the coating layers. In particular, the third term is due 
to fluctuations in the total phase the light gains when propagating within the $j$-th layer, while the fourth term is due to 
the (effective) reflectivity of the $p$-th interface (with $p=0$ indicating the coating-air interface), whose origin will be 
explained below.

\subsection{Fluctuations $\delta\phi_j$ and $\delta r_p$}
\label{sec:fluctuation}

\label{app:n_fluc}

\R{Light propagating within the coating layers are affected by the {\it photoelastic} effect, namely an {\it isothermal} fluctuation in $\delta n_j (\mathbf{x})$ (note here that $\mathbf{x}$ is a 3-D vector) due to 
fluctuating Brownian stresses exerted onto the coating materials. Assuming {\it isotropy} of the coating materials, we 
can write
\begin{eqnarray}
\label{eq:dn}
\delta n_j (\mathbf{x}) &= &\beta_j^L   S_{zz}(\mathbf{ x})   +
\beta_j^T   \left[S_{xx}(\mathbf{x})+S_{yy}(\mathbf{x})\right] 
\end{eqnarray}
with \begin{equation}
\beta_j^L \equiv \left(\frac{\partial  n_j}{\partial \log l} \right)_{A_j}\,,\quad
\beta_j^T \equiv \left(\frac{\partial  n_j}{\partial \log A}\right)_{l_j}
\end{equation} 
Here $L$ stands for longitudinal, and $T$ stands for transverse, and the subscript $A_j$ 
and $l_j$ indicate fixing 
transverse area and longitudinal length, respectively.  We have also used the usual strain definition 
\begin{equation}
S_{ij} \equiv 
\frac{1}{2}\left[\frac{\partial u_i}{\partial x_j} + \frac{\partial u_j}{\partial x_i}\right]
\end{equation}
where $u_i(\mathbf{x})$, $i=1,2,3$ are components of the displacement vector of the mass 
element at position $\mathbf{x}$.  Please refer to Appendix~\ref{app:elasticity} 
for more details in defining the elasticity 
quantities, and Appendix~\ref{photoelastic} for more details on the photo elastic effect. }

\R{We note that in Eq.~\eqref{eq:dn} $S_{zz}$ is the fractional increase in length (i.e., linear expansion) in the longitudinal 
direction, while $S_{xx}+ S_{yy}$ 
is the fractional increase in the transverse area. According to Appendix~\ref{app:transverse}, we can ignore the second term representing area fluctuations in 
Eq.~\eqref{eq:dn}  when the beam spot size  is much larger than the coating thickness. 
In this case, we write $\beta_j$ in place for $\beta_j^L$, whose value can be expressed in 
terms of a particular component of the photo elastic tensor, see Eq.~\eqref{betaL}. } 

\R{As we discuss in Appendix~\ref{app:inf:layer},  the (surviving)  first term of Eq.~\eqref{eq:dn} causes two effects for light propagating along each direction (i.e., $+z$ and $-z$): it adds an additional phase shift onto the light, and it back-scatters a fraction of the light into the opposite direction. As we show in Appendix~\ref{app:stack} [c.f. Eqs.~\eqref{dr}--\eqref{dphijapp}], theses effects can be described by modifying the phase shift $\delta\phi_j$ of each coating layer and the  reflectivity $\delta r_j$ of interface:}
\begin{eqnarray}
\label{dphij}
\delta \phi_j &=& k_0\Big[ (n_j+\beta_j)\delta l_j - \frac{1-r_j^2}{2r_j}\beta_j \delta l_j^c 
\nonumber\\
&&\qquad\qquad\qquad\;\;\;
+\frac{1+r_{j-1}^2}{2r_{j-1}}\beta_{j-1} \delta l_{j-1}^c\Big] \,,\\
\label{drj}
\delta r_j &=& k_0 t_j^2 \beta_j \delta l_j^s\,.
\end{eqnarray}
Here we have defined  
\begin{eqnarray}
\label{eqljc}
\delta l_j^c &=& -\int_0^{l_j} S_{zz}(z_{j+1}+z) \cos (2k_0 n_j z) dz \\
\label{eqljs}
\delta l_j^s &=& -\int_0^{l_j} S_{zz}(z_{j+1}+z) \sin (2k_0 n_j z) dz
\end{eqnarray}
%
for $j\ge 1$, $\delta l^s_0 = \delta l^c_0=0$, and
\begin{equation}
z_{j} \equiv \sum_{n=j}^{N} l_n\,.
\end{equation}
marks the $z$-coordinate of the top surface of the $j$-th layer. We can also write
\begin{equation}
\delta l_j = \int_0^{l_j} S_{zz}(z_{j+1}+z)dz\,.
\end{equation}
 Note that
\begin{equation}
\begin{array}{c}
\mbox{total coating} \\ \mbox{thickness}
\end{array} = z_1 > z_2 > \ldots > z_{N+1}=0
\end{equation}
Note that $\delta r_j$, as well as the last two terms in $\delta\phi_j$ \R{are due to back-scattering, and} have not been considered by 
previous authors. 

Inserting Eqs.~\eqref{dphij}, \eqref{drj}  into Eq.~\eqref{coth}, we obtain:
\begin{equation}
\label{eq:xi}
\xi(\vec x) -i\zeta (\vec x ) =   -z_s(\vec x) -\sum_{j=1}^N\int_{z_{j+1}}^{z_{j}} \left[1+\frac{i\epsilon_j(z)}{2}\right]u_{zz}(\vec x,z) dz
\end{equation}
where
\begin{eqnarray}
\label{epsilon}
\epsilon_j (z)&=& (n_j+\beta_j)\frac{\partial\log\rho}{\partial\phi_j}\nonumber\\
&-&\beta_j\bigg[\frac{1-r_j^2}{2r_j}\frac{\partial\log\rho}{\partial\phi_j}
\nonumber\\
&&\quad -\frac{1+r_j^2}{2r_j}\frac{\partial\log\rho}{\partial\phi_{j+1}}\bigg] \cos [2k_0n_j(z-z_j)]\nonumber\\
&-&t_j^2 \beta_j  \frac{\partial\log\rho}{\partial r_j} \sin[2k_0n_j(z-z_{j+1})]\,,
\end{eqnarray}
a  term that  accounts for all effects associated with light penetration.  
Here we need to formally define
\begin{equation}
\frac{\partial \log\rho}{\partial \phi_{N+1}}=0
\end{equation}
since $\phi_{N+1}$ does not really exist. 
Alternatively, we can also write formulas separately for $\xi$ and $\zeta$, using only real-valued 
quantities.  For $\xi$, we have,
\begin{eqnarray}
\label{xi:terms}
\xi(\vec x) &=& - z_s(\vec x) 
\nonumber\\
&&- \sum_{j=1}^N \left[\mathcal{T}^\xi_j \delta l_j (\vec x)+
\mathcal{T}_j^{\xi  c} \delta l_j^{c} (\vec x)+
\mathcal{T}_j^{\xi  s} \delta l_j^{s} (\vec x)\right], \quad\;
\end{eqnarray}
where 
\begin{eqnarray}
\label{Tjxi}
\mathcal{T}_j^\xi &=&1 -\frac{n_j +\beta_j}{2}\mathrm{Im}\left(
\frac{\partial \log \rho}{\partial \phi_j}\right) \\
\label{Tjc}
\mathcal{T}_j^{\xi c} &=& -\frac{\beta_j}{4} \mathrm{Im}\left(\frac{\partial \log \rho}{\partial \phi_{j}}\right)\left(\frac{1-r_j^2}{r_{j}}\right)\nonumber\\
&&+\frac{\beta_j}{4}\mathrm{Im}\left(\frac{\partial \log \rho}{\partial \phi_{j+1}}\right)\left(\frac{1+r_{j}^2}{r_{j}}\right)\\
\label{Tjs}
\mathcal{T}_j^{\xi s} &=& -\frac{\beta_j t_{j}^{2}}{2}\mathrm{Im}\left(\frac{\partial \log\rho}{\partial r_{j}}\right) 
\end{eqnarray}
are transfer functions from the various $\delta l$'s to the displacement-equivalent thermal noise. 
For $\zeta$, we have
\begin{equation}
\zeta(\vec x) = \sum_{j=1}  \left[
\mathcal{T}_{j}^\zeta\delta l_{j}(\vec x) 
+
\mathcal{T}_{j}^{\zeta c}\delta l_{j}^{c}(\vec x) 
+
\mathcal{T}_{j}^{\zeta s}\delta l_{j}^{s}(\vec x) 
\right]
\end{equation}
where
\begin{eqnarray}
\label{Sjxi}
\mathcal{T}_{j}^\zeta &=& \frac{n_j +\beta_j}{2}\mathrm{Re}\left(
\frac{\partial \log \rho}{\partial \phi_j}\right)\\
\label{Sjc}
\mathcal{T}_{j}^{\zeta c} &=& \frac{\beta_j}{4} \mathrm{Re}\left(\frac{\partial \log \rho}{\partial \phi_{j}}\right)\left(\frac{1-r_j^2}{r_{j}}\right)\nonumber\\
&-&\frac{\beta_j}{4}\mathrm{Re}\left(\frac{\partial \log \rho}{\partial \phi_{j+1}}\right)\left(\frac{1+r_{j}^2}{r_{j}}\right)\\
\label{Sjs}
\mathcal{T}_{j}^{\zeta s} &=&\frac{\beta_{j} t_{j}^{2}}{2}\mathrm{Re}\left(\frac{\partial \log\rho}{\partial r_{j}}\right)
\end{eqnarray}
%
%
%
%

Although for an arbitrary stack of dielectrics, $\zeta$ is comparable to the part of $\xi$ 
[c.f.~Eq.~\eqref{epsilon}] that 
involves light penetration into the layers.  In practice, for highly reflective stacks, the real 
parts of 
$\partial\log\rho/\partial\phi_j$ and $\partial\log\rho/\partial r_j$ all turn out to be 
small, and therefore fluctuations in $\zeta$ should be much less than fluctuations in $\xi$.

%

\subsection{Mode selection for phase noise}
\label{phasethermalnoise}
So far we have calculated phase and amplitude noise as functions of location $\vec x$ on the mirror surface. However, there is only one displacement noise that the light will sense.  In this and the next subsection, \R{show} how $\xi (\vec x)$ and $\zeta (\vec x)$ should be converted into \R{measurement noise}.  In doing so, we recognize that only one spatial optical mode is \R{injected} on resonance in the optical cavity, and this mode has a complex amplitude of $ u_0(\vec x)$ at the mirror surface. Now suppose we have $u_{\rm in} = u_0(\vec x)$ incident on the mirror surface, we will then have $u_{\rm out}(\vec x) = \rho_{\rm tot}(\vec x) u_0(\vec x)$, which contains not only the resonant mode, but also other modes, which do not resonate in the cavity.

Let us select only the component of $u_{\rm out}(\vec x)$ \R{resonates}, then we have a complex reflectivity of
\begin{equation}
\label{average}
\bar \rho =\frac{\int  u_{0}^*(\vec x) u_{\rm out}(\vec x)  d^2\vec x}{\int u_{0}^* u_{0} d\vec x}  = \frac{\int \rho_{\rm tot}(\vec x) I(\vec x) d^2\vec x}{\int I(\vec x) d^2\vec x}\,,
\end{equation}
specifically for the resonant mode, and hence independent of $\vec x$.  Here we have defined $I(\vec x) \equiv |u_0(\vec x)|^2$. Note that the bar on top of $\bar\rho$ represents averaging over the phase front, instead of averaging over time. 

\R{Now, inserting Eq.~\eqref{rho_re_im} as definitions for $\xi(\vec x)$ and $\zeta(\vec x)$ into Eq.~\eqref{average}, we obtain the fluctuating part of $\bar\rho$
\begin{equation}
\label{drhobar}
\frac{\delta\bar\rho}{\bar\rho} = 2 i k_0 (\bar \xi - i \bar\zeta)\,,
\end{equation}
where 
\begin{equation}
\label{kaiave}
\bar \xi \equiv \frac{ \int  \xi(\vec x) I(\vec x) d^2\vec x}{  \int I(\vec x) d^2\vec x}\,,\quad \bar\zeta\equiv \frac{\int \zeta(\vec x) I(\vec x) d^2\vec x}{\int  I(\vec x) d^2\vec x}
\,.
\end{equation}
Note that $2ik_0\bar\xi$ is the additional phase gained by the returning light, while $2 k_0 \bar\zeta$ is the relative change in amplitude [see discussions in Sec.~\ref{subsec:pham}].  Focusing first on $\bar\xi$, we note that this creates the same phase as that gained by the reflected light if the mirror does not deform but instead moves by $\bar\xi$. In this way, $\bar\xi$ is an error in our measurement of the mirror's displacement. }

\subsection{Conversion of Amplitude Noise into Displacement}
\label{subsec:amp}
The amplitude thermal noise produces spurious GW signal by modulating the radiation pressure acting on the mirror, which in turn drives spurious \R{mirror} motion. Let us first consider a single-bounce scenario, in which an incoming beam with intensity profile $I(\vec x)$, unaffected by thermal noise, is reflected with an intensity profile $I(\vec x)+\delta I(\vec x)$, with $\delta I(\vec x)$ induced by amplitude thermal noise.
\R{In this case,  the mirror feels a thermal-noise-induced  recoil force of
\begin{equation}
F_{\rm th}^{\rm single} =\int \frac{\delta I(\vec x)}{c} d^2\vec x \,.
\end{equation}
Using Eqs.~\eqref{ampth} and \eqref{kaiave}, we obtain
\begin{equation}
F_{\rm th}^{\rm single} =\frac{4 I_0 k_0}{c}\bar\zeta
\end{equation}}
with $I_0$ the power incident on the mirror.   If the mirror is within a cavity, then we need to consider both the \R{increase in} the circulating power (which we denote by $I_c$) with respect to the input power, and the coherent build-up of amplitude modulation within the cavity.  We also note that now both the incident and reflected beam contains amplitude modulation, and that \R{we must also consider the effect of this amplitude modulation on the input mirror}. 

If we restrict ourselves to a single \R{optical} cavity \R{on} resonance, then the force thermal noise \R{below the cavity bandwidth} is given by   
\begin{equation}
F_{\rm th}^{\rm cav} = \frac{16k_0  I_c}{c\sqrt{T_{\rm i}}}  \bar\zeta 
\end{equation}
Here $I_c$ is the circulating power in the arm cavity.  Suppose both input and end mirrors have the same mass $M$, then the spectrum of cavity length modulation driven by the amplitude thermal noise is given by 
\begin{equation}
\label{ampconv}
\sqrt{S_{\rm th}^{\rm amp}} =\frac{2}{m\Omega^2}\sqrt{S_{F_{\rm th}^{\rm cav}}}= \frac{32\omega_0 I_c}{m\Omega^2 c^2\sqrt{T_{\rm i}}}\sqrt{S_{\bar\zeta}} 
\end{equation}
Note that $\bar \zeta$ has the units of displacement, and therefore the pre-factor in 
front of $\sqrt{S_{\bar\zeta}}$ in Eq.~\eqref{ampconv} is a dimensionless conversion factor 
from $\bar\zeta$ to displacement noise.  For Advanced LIGO, this cannot be completely 
dismissed at this stage, because 
\begin{equation}
\frac{32\omega_0 I_c}{m \Omega^2 c^2\sqrt{T_{\rm i}}} = 18 \cdot\frac{I_c}{800\,{\rm kW}}\cdot\frac{40\,{\rm kg}}{m}\cdot
\left[\frac{ 10\,{\rm Hz}}{\Omega/(2\pi)}\right]^2\sqrt{\frac{0.03}{T_{\rm i}}} \end{equation}
Nevertheless, as we will show in Sec.~\ref{subsec:penetration}, the minor amplification factor 
here is not enough in making amplitude noise significant, because $\zeta$ tend to be much 
less than $\xi$, for the coatings we consider.

\section{Thermal noise assuming no light penetration into the coating}
\label{sec:nopenetration}

In this section, we compute the coating Brownian noise assuming that the incident light 
does not penetrate into the coating. This means light is promptly reflected at the coating-air 
interface, and therefore we should only keep the first two terms on the right-hand side of 
Eq.~(\ref{coth}), which leads to $\zeta=0$. We therefore 
\R{consider only coating phase noise} $\xi$, in particular its weights average throughout the mirror surface, $\bar\xi$, see Eq.~\eqref{kaiave}.

\subsection{The Fluctuation-Dissipation Theorem}
\label{sec:FDT}
The Fluctuation-Dissipation Theorem relates the near-equilibrium thermal noise spectrum of a 
generalized coordinate $q$ to the rate of dissipation in the system when a generalized force acts 
directly on this coordinate.  More specifically, the thermal noise spectrum of $q$ at temperature 
$T$ is given by~\cite{Callen:FDT}
\begin{eqnarray}
\label{Sxf}
S_q(f)=\frac{k_B T}{\pi^2 f^2}\mathrm{Re}[Z(f)]
\end{eqnarray}
where $f$ is frequency, $Z(f)$ is the mechanical impedance (inverse of admittance), or
\begin{equation}
Z(f) = -2\pi i f q(f)/F(f)
\end{equation}
Alternatively, imagining a sinusoidal force
\begin{equation}
F(t) =F_0\cos(2\pi f t)
\end{equation}
 with amplitude $F_0$ acting directly on $q$, Eq.~(\ref{Sxf}) can also be written as
\begin{eqnarray}
\label{SxfU}
S_x(f)=\frac{4k_B T}{\pi f}\frac{W_{\rm diss}}{F_0^2}= \frac{4k_B T}{\pi f}\frac{U}{F_0^2}\phi
\end{eqnarray}
where $W_{\rm diss}$ is the energy dissipated per cycle of oscillation divided by $2\pi$ (in other words, 
$W_{\rm diss}$ is the average energy loss per radian), 
$U$ is the peak of the stored energy in the system, and $\phi$ is the loss angle, defined by
\begin{equation}
\phi = \mathrm{Re}[Z(f)]/\mathrm{Im}[Z(f)]
\end{equation}

It is important to note that $\phi$ is in general frequency dependent.  However, for an elastic body, 
if the frequency is low enough (much below the first eigenfrequency), then $U$ can be computed 
\R{using the quasi-static approximation}, because it is equal to the elastic energy stored in the equilibrium 
configuration when \R{a constant} $F_0$ is applied to the system.

\subsection{Mechanical Energy Dissipations in Elastic Media}
\label{subsec:med}

It is straightforward to apply Eq.~(\ref{SxfU}) to calculate the thermal noise component due to fluctuation of the position of the coating-air interface --- the weighted average [c.f.~Eq.~\eqref{average}] of the first two terms of Eq.~(\ref{coth}).  This can be obtained by applying a force $F$ with a pressure profile proportional to $I(\vec x)$ on to the mirror surface (coating-air interface).   In this case, elastic energy can be divided into bulk energy $U_B$ and shear energy $U_S$ [Chapter I of Ref.~\cite{Landau}], with
\begin{equation}
U_{\rm coating} =U_B + U_S =\int_{\rm coating} \left(\frac{K}{2 } \Theta^2 +  \mu \Sigma_{ij} \Sigma_{ij} \right) dV\,,
\end{equation}
where $\Theta$ is the expansion, and $\Sigma_{ij}$ is the shear tensor (see Appendix~\ref{app:elasticity} for details).   \R{If we give small imaginary parts to $K$ and $\mu$, writing}
\begin{equation}
\tilde K = K (1+i\phi_{\rm B})\,,\quad \tilde \mu = \mu(1+i\phi_{\rm S})
\end{equation}
then $W_{\rm diss}$ can be written as
\begin{eqnarray}
\label{eq:wdiss}
W_{\rm diss} =\phi_B U_{B} +\phi_{S} U_{S}
\end{eqnarray}
\R{Here have introduced the loss angles {\bf $\phi_B$} and {\bf $\phi_S$}, which are associated with the dissipation of
expansion energy density and the shear energy density, respectively. Note that our way of characterizing loss} differs from previous work by Harry, et.\ al.~\cite{harry}, because for isotropic 
materials, {\bf $\phi_B$} and {\bf $\phi_S$} are the \R{two fundamentally independent loss angles that characterize the dissipation of bulk and shear elastic energy;} were 
we to \R{literally} adopt $\phi_\perp$ and $\phi_\parallel$ as done in Ref.~\cite{harry}, there would be \R{modes} of external 
driving that lead to negative dissipative energy, as shown explicitly in Appendix~\ref{App:lossangle}. 

\R{Once we have introduced $\phi_{\rm B}$ and $\phi_{\rm S}$, other elastic moduli also gain small imaginary parts correspondingly. For example, for the most widely used Young's modulus and Poisson ratio, because}
\begin{eqnarray}
K=\frac{Y}{3(1-2\sigma)}, \quad \mu=\frac{Y}{2(1+\sigma)}
\end{eqnarray}
we can write
\begin{equation}
\tilde Y = Y(1+i\phi_Y)
\end{equation}
with
\begin{equation}
\phi_Y = \frac{(1-2\sigma)\phi_B+2 (1+\sigma)\phi_S}{3}
\end{equation}
and
\begin{equation}
\tilde \sigma = \sigma+\frac{i}{3}(1-2\sigma)(1+\sigma)(\phi_B-\phi_S)\,.
\end{equation}
Since $-1 < \sigma <1/2$, we have $(1-2\sigma)(1+\sigma) >0$, therefore $\tilde\sigma$ has a positive imaginary part as $\phi_B$ is greater than $\phi_S$, and vice versa. \R{To understand the physical meaning of the imaginary part of Poisson ratio, one has to realize that Young's modulus and Poisson ratio together describe the elastic response of a rod.  Suppose we apply an oscillatory} tension uniformly along a rod at a very low frequency, whether the area of the rod leads or lags the length of the rod depends on the relative magnitudes of the bulk and shear loss angles. In the situation when the two loss angles $\phi_B$ and $\phi_S$ are equal to each other, the Poisson's ratio is real, and we only need to deal with one loss angle $\phi_Y$ --- although there is reason to assume the equality of these two angles.   

\R{If the coating material is made into the shape of a one-dimensional rod, and if we only consider its elongational, bending or torsional modes, then the Young's modulus is the appropriate elastic modulus associated with these modes, and $\phi_Y$ is the appropriate loss angle to apply.  However, this is not directly relevant for coating thermal noise. An elastic modulus that will actually} prove useful is that of the {\it two-dimensional (2-D) flexural rigidity} of a thin plate made from the coating material, 
\begin{equation}
D =\frac{Y h^3}{12(1-\sigma^2)} =|D|(1+i\phi_D)
\end{equation}
where $h$ is the thickness of the plate, with
\begin{equation}
\label{eqphiD}
\phi_D  =   \frac{(1-\sigma-2\sigma^2)\phi_B +2(1-\sigma+\sigma^2)\phi_S}{3(1-\sigma)}\,.
\end{equation}
As we shall see in Sec.~\ref{subsec:bend}, this $D$ is most easily measured through the quality factor of drum modes of a thinly coated sample --- although this will not turn out to be the combination of loss angle that appear in the thermal noise.

\subsection{Thermal Noise of a Mirror Coated with one Thin Layer}
\label{subsub:tnc}

In the case where the coating thickness is much less than the size of the \R{mirror substrate}  and the beam spot size,  the elastic deformation of the substrate is not affected by the presence of the coating. As a consequence, if  we include the elastic energy stored in the substrate $U_{\rm sub}$ with loss angle $\phi_{\rm sub}$, we can write
\begin{eqnarray}
W_{\rm diss} &=&\phi_{\rm sub} U_{\rm sub} + \phi_B U_{B} +\phi_{S} U_S \nonumber \\
&  \approx & \left[\phi_{\rm sub}  + \phi_B  \frac{U_{B}}{U_{\rm sub}} +\phi_S \frac{U_S}{U_{\rm sub}}\right] U_{\rm sub}
\end{eqnarray}
With the assumption of thin coating and half-infinite substrate, the total strain energy stored in the sample can be considered as $ U_{\rm sub}$. In such a way the coating adds on to substrate loss angle as  additional, effective angles
\begin{equation}
\label{eq:totalangle}
\phi_{\rm coated}=\phi_{\rm sub}+\frac{U_{B}}{U_{\rm sub}}\phi_B+\frac{U_S }{U_{\rm sub}}\phi_S
\end{equation}
Note that when the total coating thickness $l $ is much less than beam spot size $w_0$, we have $U_B/U_{\rm sub}\sim U_S/U_{\rm sub} \sim l/w_0 \ll 1$.  Unfortunately, however,  $\phi_B$ and $\phi_S$ are found to be so much larger than the substrate loss angle $\phi_{\rm sub}$ that in practice coating thermal noise still dominates over substrate thermal noise.

Now suppose we would like to measure a weighted average of the position of the mirror surface,
\begin{equation}
\label{qweight}
q = \bar\xi = \int d^2 \vec x\, w(\vec x) z(\vec x)
\end{equation}
with [Cf.~Eq.~\eqref{kaiave}]
\begin{equation}
w(\vec x) = \frac{I(\vec x)}{\int I(\vec x) d^2\vec x}
\end{equation}
and $z(\vec x)$ the position of the coating-air interface at transverse location $\vec x$.

According to Sec.~\ref{sec:FDT}, we need to apply a pressure profile of
\begin{equation}
f(\vec x) = F_0 w(\vec x)
\end{equation}
onto the upper surface of the coating, which we shall also refer to as the coating-air interface.  Straightforward calculations give

\begin{eqnarray}
\label{eq:Utheta}
\frac{U_{B}}{F_0^2}&=&\frac{(1-2\sigma_c) l}{3 }
\bigg[ \frac{Y_c}{Y_s^2}\frac{(1-2\sigma_s)^2(1+\sigma_s)^2}{(1-\sigma_c)^2} \nonumber \\
&&\qquad\qquad \;\;\,+\frac{1}{Y_s}\frac{2(1-2\sigma_s)(1+\sigma_s)(1+\sigma_c)}{(1-\sigma_c)^2} \nonumber \\
&&\qquad\qquad \;\; \,+\frac{1}{Y_c}\frac{(1+\sigma_c)^2}{(1-\sigma_c)^2}\bigg] \int w^2(\vec x) d^2\vec x\\
 \frac{U_{S}}{F_0^2}&=&\frac{2 l}{3 }
 \bigg[
 \frac{Y_c}{Y_s^2}\frac{(1-\sigma_c+\sigma_c^2)(1+\sigma_s)^2(1-2\sigma_s)^2}{(1-\sigma_c)^2(1+\sigma_c)}
\nonumber\\
\label{eq:Usigma}
 &&\quad\;-\frac{(1+\sigma_c)(1-2\sigma_c)(1-2\sigma_s)(1+\sigma_s)}{Y_s(1-\sigma_c)^2}\nonumber \\
 &&\quad\;
+ \frac{(1-2\sigma_c)^2(1+\sigma_c)}{Y_c(1-\sigma_c)^2} \bigg]\int w^2(\vec x)d^2\vec x
\end{eqnarray}
Here $l$ is coating thickness; for Young's modulus $Y$ and Poisson's ratio $\sigma$,  substrates $c$ and $s$ represent coating and substrate, respectively.  Directly following Eqs.~\eqref{SxfU} and \eqref{eq:wdiss} will give rise to a  noise spectrum of
\begin{equation}
\label{sxinonp}
S_{\bar \xi} = \frac{4k_B T}{\pi f}\left[\phi_B \frac{U_{B}}{F_0^2} + \phi_S \frac{U_{S}}{F_0^2}\right]
\end{equation}
where $U_{B}/{F_0^2}$ and $U_{S}/{F_0^2}$ are given by Eqs.~\eqref{eq:Utheta} and \eqref{eq:Usigma} respectively. 

Here we can define
\begin{equation}
\int w^2(\vec x) d^2 \vec x = \frac{\int d^2 \vec x I^2(\vec x)}{\left[\int d^2 \vec x I(\vec x)\right]^2} \equiv  \frac{1}{\mathcal{A}_{\rm eff}}
\end{equation}
as the inverse of an {\it effective beam area}. Therefore noise power in $q$ is proportional to coating thickness and inversely proportional to beam area.  In particular, for a Gaussian beam with
\begin{equation}
\label{gaussianbeam}
I(\vec x) \propto \exp\left(-\frac{\vec 2 x^2}{w_0^2}\right)
\end{equation}
the effective area is $\mathcal{A}_{\rm eff} =  \pi w_0^2$.

Let us compare our results to previous calculations using $\phi_\perp$ and $\phi_\parallel$. As it turns out, if we assume $\phi_S=\phi_B$, then formulas for thermal noise agree with Eq. (22) in Ref.~\cite{harry}. To illustrate the different roles now played by $\phi_B$ and $\phi_S$, let us take the \R{very simple} case of  $Y=Y_c=Y_s$ and $\sigma=\sigma_c=\sigma_s$, where
\begin{eqnarray}
&&\frac{\delta U_{B}}{F_0^2}= \frac{4l}{3Y \mathcal{A}_{\rm eff}}(1+\sigma)^2(1-2\sigma) \\
&&\frac{\delta U_{S}}{F_0^2} =\frac{2l}{3Y \mathcal{A}_{\rm eff}}(1+\sigma)(1-2\sigma)^2 
\end{eqnarray}
Using Eq.~\eqref{sxinonp}, we can get the power spectral density of the single layer non-penetration coating thermal noise as
\begin{eqnarray}
\label{eq:tn:simple}
&&S_{\bar \xi}(f) \nonumber\\
\!\!\!&=&\!\!\!\frac{8 k_B T (1-\sigma-2\sigma^2)l}{3\pi f  Y \mathcal{A}_{\rm eff} } [2(1+\sigma)\phi_B+(1-2\sigma)\phi_S].\;\;\quad
\end{eqnarray}
From Eq.~\eqref{eq:tn:simple}, we can  see that the bulk loss and shear loss contribute differently to the total noise. More importantly, \R{at least in this very simple case of $Y_{c}=Y_s$,} the combination of $\phi_B$ and $\phi_S$, approximately $2\phi_B +\phi_S$, that enters the thermal noise \R{apparently differs significantly} from the combination  $\phi_{\rm tot} \approx \phi_B+2\phi_S$, \R{which has been} measured by ring-down experiments that have been performed so far~\cite{ringdown1,ringdown2,ringdown3} --- as we will see in Eq.~\eqref{eq:ringdownphi} and will be discussed in detail in the rest of Sec.~\ref{sec:measurement}.

\subsection{\R{Discussions on the correlation structure of thermal noise}}
\label{subsubsec:stn}

Before proceeding to more detailed calculations of Brownian noise that involve light penetrating into the coating layers, we would like to gain more insight about thermal noise by inspecting our existing expressions of coating thermal noise [Eqs.~\eqref{eq:Utheta}--\eqref{sxinonp}] more carefully.  We note that 
\begin{equation}
S_{\bar\xi}  \propto l \int w^2(\vec x)d^2 \vec x.
\end{equation}
where the coefficient of proportionality depends only on  material property.  From such a dependence on  coating and beam geometries, we deduce that (i) each point on the coating-air interface fluctuates along the $z$ direction independently, and (ii) materials at different $z$'s within the coating also contribute independently to coating thermal noise.  These observations will be confirmed mathematically in the next section. 

Finally, within the coefficient of proportionality [Cf.~Eqs.~\eqref{eq:Utheta} and \eqref{eq:Usigma}], we found three types of dependence on the Young's moduli of the coating and substrate materias: terms proportional to $1/Y_c$ are expected to arise from fluctuations in coating thickness,  terms proportional to $Y_c/Y_s^2$ can be interpreted as arising from coating thermal stresses driving the substrate-coating interface, while terms proportional to $1/Y_s$ are therefore interpreted as correlations between the above two types of noise.

\section{Cross Spectra of Thermal Noise components} 
\label{sec:spectra}

\R{In this section, we compute the cross spectra of each component of coating thermal noise, and assemble the formula for the spectral density of the total noise.  Specifically, in Sec.~\ref{subsec:coat-thick}, we compute the cross spectra of the thickness fluctuations any two  uniform sublayers of the coating, and obtain the cross spectrum of $S_{zz}$;   in Sec.~\ref{subsec:csint}, we compute the cross spectra involving height fluctuation $z_s$ of the coating-substrate interface, i.e., $S_{S_{zz}z_s}$ and $S_{z_s z_s}$;  in Sec.~\ref{subsec:anatomy}, we dissect the above results and analyze the separate roles of bulk and shear fluctuations;  in Sec.~\ref{subsec:fullformula}, we write down the full formula of coating thermal noise. }

\subsection{Coating-Thickness Fluctuations}
\label{subsec:coat-thick}

\begin{figure}[h]
\centering
\includegraphics[width=0.45\textwidth]{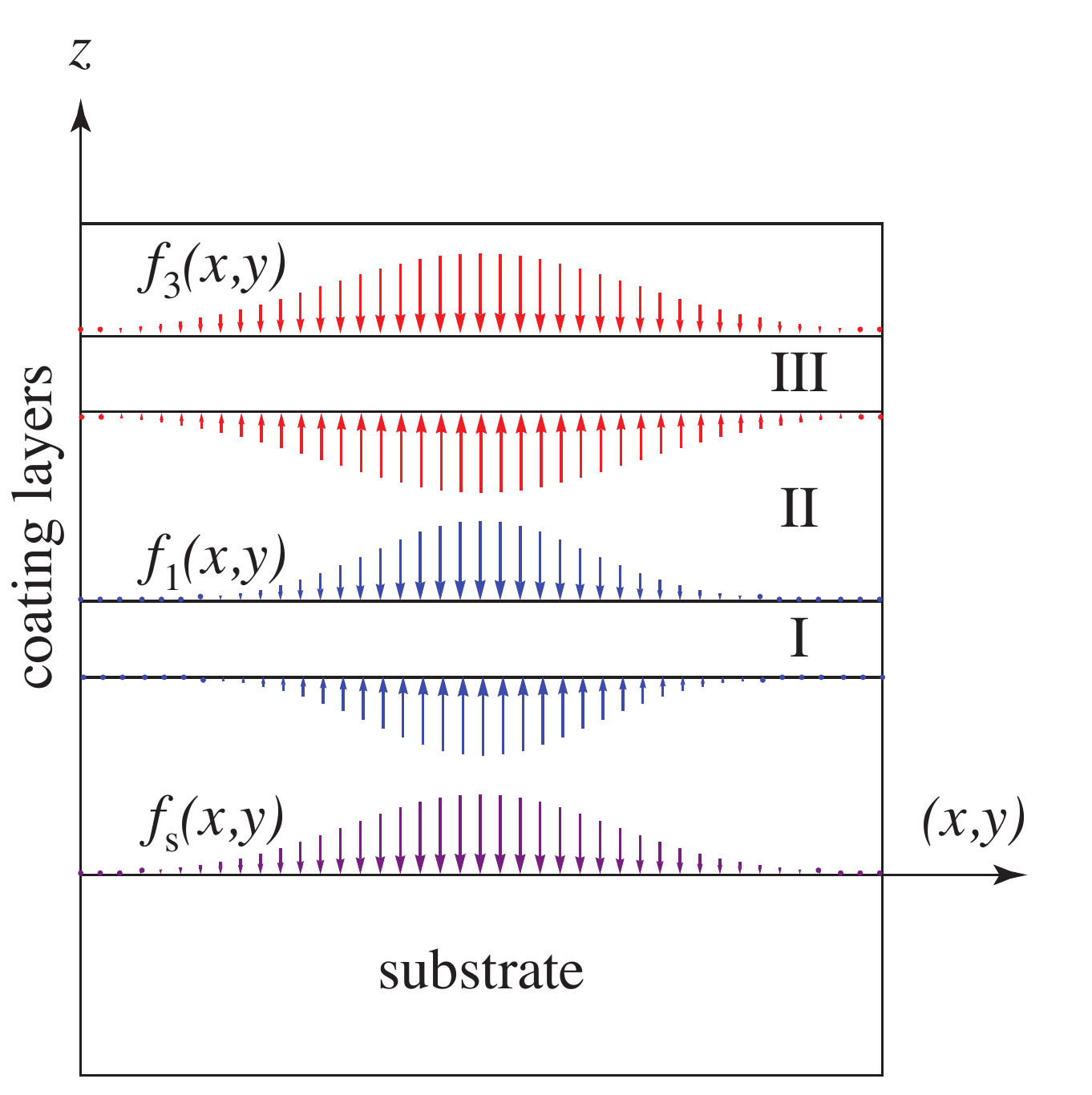}
\caption{Illustrations of forces applied onto various interfaces within the coating.  Each of Layers I and III in the coating are assumed to be uniform (but they might each contain a different kind of material); region II denotes the entire gap between them, which may well contain many different dielectric layers.    A pair of force distribution $f_1$ ($f_3$) in opposite directions is exerted on opposite sides of Layer I (III), while  $f_s$ is exerted on the coating-substrate interface. The three distributions may well have different profiles (as also illustrated in the figure).
 \label{force}}
\end{figure}

Let us start by calculating thickness fluctuations of individual layers and correlations among them. Following Levin's approach, we imagine applying two pairs of opposite pressure,
\begin{eqnarray}
f_1(\vec x) = F_0 w_1(\vec x), \quad f_3(\vec x) &=&F_0 w_3(\vec x)
\end{eqnarray} in the $z$ direction on layer {\Rmnum 1}  and layer {\Rmnum 3}, as shown in Fig.~\ref{force}, with thickness of $l_1$ and $l_3$, respectively. Here $w_1(\vec x)$ and $w_3(\vec x)$, like the $w(\vec x)$ used in Eq.~\eqref{qweight}, provides the shape of the pressure profiles. 

We assume that within each of I and III, there is only one type of material, yet there could be arbitrary number of different material sub layers in II.   As it will turn out, the precise locations of layers I and III along the $z$ direction does not affect the result, {\it as long as they do not overlap}, or in other words, layer II has non-zero (positive) thickness.

Throughout this paper, we shall assume that the beam spot size is much less than the radius of the mirror, so that we can make the approximation that the mirror surface is an infinite two-dimensional plane.  In this case, we perform a spatial Fourier transformation for the applied pressure,
\ba
\tilde f_j(\vec k)=\int e^{i \vec k \cdot \vec x}f_j(\vec x)\, d^2 \vec x = F_0 \tilde w_j(\vec k) \,,\; j=1,3,
\ea
and carry out our calculations for strain and stress distributions in the coating-substrate system in the Fourier domain. 
%

We further assume that the coating thickness is much less than the beam spot size, which is inverse the maximum spatial frequency contained in $\tilde w_{1,3}$. This means we only need to consider $\vec k$'s with $|\vec k| l \ll 1$, with $l$ the total coating thickness.   According to calculations of Appendix~\ref{app:elasticity}, non-zero components of the stress and strain tensors in Layers I and III  are found to be (in the spatial Fourier domain)
\be
\label{eq:T1}
\tilde T^{\rm I}_{xx}=\tilde T^{\rm I}_{yy} =\frac{\sigma_1 \tilde w_1}{1-\sigma_1}F_0 \,,\quad \tilde T^{\rm I}_{zz} = \tilde w_1 F_0\,,
\ee
\be
\label{eq:S1}
\tilde S^{\rm I}_{zz} = -\frac{(1-2\sigma_1)(1+\sigma_1) \tilde w_1 }{Y_1(1-\sigma_1)}F_0\,,
\ee
and
\be
\tilde T^{\rm III}_{xx}=\tilde T^{\rm III}_{yy} =\frac{\sigma_3 \tilde w_3}{1-\sigma_3}F_0 \,,\quad \tilde T^{\rm III}_{zz} = \tilde w_3 F_0\,,
\ee
\be
\tilde S^{\rm III}_{zz} = -\frac{(1-2\sigma_3)(1+\sigma_3) \tilde w_3 }{Y_3(1-\sigma_3)}F_0\,,
\ee
respectively.  

\R{Note that deformations within layer I only depends on  $\tilde w_1$ (not $\tilde w_3$), while deformations within layer III only depends on $\tilde w_3$ (not $\tilde w_1$) --- while regions outside these layers are found to have vanishing strain and stress.  This means we can treat deformations caused by each pair of forces independently, as long as layer I and layer III do not overlap. The deformations are also independent from the thickness of the layers.  The vanishing of deformations outside these layers means that when we introduce additional pairs of opposite forces, the new deformations introduced will be constrained within those new layers --- as long as those new layers do not overlap with existing ones.
This independence originates from the linearity of elastic response, and the fact that coating strains induced by force applied on a single surface within the coating, as discussed in Appendix.~\ref{app:elasticity}, does not depend on distance away from that surface, as seen in Eqs.~\eqref{sxxa}--\eqref{szzb}.  The situation here is analogous to the electric field generated by several pairs of oppositely-charged infinite parallel planes.   }

In terms of thermal noise, such a distribution of elastic deformations corresponds to a dissipation energy that consists of two independent terms, each corresponding to one layer and proportional to its thickness:
\begin{equation}
\frac{W_{\rm diss}}{F_0^2} = W_{11} l_1\int w_1^2 d^2\vec x + W_{33} l_3\int w_3^2 d^2\vec x 
\end{equation} 
Here we have defined, for $J=1,3$:
\begin{equation}
\label{wjj}
W_{jj} \equiv  \frac{(1-2\sigma_j)(1+\sigma_j)}{3(1-\sigma_j)^2 Y_j } \left[\frac{1+\sigma_j}{2}\phi_B^j+(1-2\sigma_j)\phi_S^j\right]\,.
\end{equation}
This means the fluctuation of
\begin{equation}
q \equiv \int \left[w_1 (\vec x) \delta l_1(\vec x) + w_3 (\vec x) \delta l_3(\vec x)\right] d^2\vec x
\end{equation}
is given by
\begin{equation}
S_q = \frac{4k_B T}{\pi f}\sum_{J=1,3} \left[W_{jj} l_j \int w_j^2(\vec x) d^2\vec x\right]
\end{equation}
The absence of cross terms here means that fluctuations in $\delta l_1(\vec x)$ and $\delta l_3(\vec x')$ are uncorrelated --- and hence statistically independent.  Furthermore, within each layer, in the same spirit as the discussions in Sec.~\ref{subsubsec:stn}, the particular form of dependence on $l_j$ and $w_j(\vec x)$ indicates that $S_{zz}$  fluctuations at different 3-D locations (within this layer) are all uncorrelated and have the same spectrum.  In this way, we obtain the cross spectral density of $S_{zz}$ at two arbitrary 3-D locations within the coating: 
\begin{equation}
\label{sllx}
S^{ij}_{S_{zz}  S_{zz}}(\vec{x},z;\vec{x}',z')
=\frac{4k_BT}{\pi f}\delta_{ij}\delta^{(2)}(\vec{x} -\vec{ x}')\delta(z-z')  W_{jj} 
\end{equation}
Here we have assumed that $(\vec x,z)$ belongs to layer $i$, while $(\vec x,z)$ belongs to layer $j$.  (The association to layers helps to identify the material property to be used in $W_{jj}$.)

\subsection{Fluctuations of Coating-Substrate Interface and their correlations with coating thickness}
\label{subsec:csint}

To investigate the correlation between height of the coating-substrate thickness, $z_s(\vec x)$ and the thickness of each coating layer, $\delta l_j(\vec x)$,  we apply a pair of pressures $f_1(\vec x) = F_0 w_1(\vec x)$ at opposite sides of Layer {\Rmnum 1}, and force $f_s(x,y)=F_0 w_s(\vec x)$ onto the coating-substrate interface (along the $-z$ direction), as shown in Fig. 1. The same strain and stress as in Eqs.~\eqref{eq:T1} and \eqref{eq:S1} are driven by $\tilde f_1$, which are only non-vanishing within layer I.   On the other hand, $\tilde f_s$ drives uniform strain and stress over the entire coating, with non-vanishing components of stress and strain given by,
\begin{equation}
\label{Tijs}
\| \tilde T_{ij} \|  = \frac{\tilde w_s (1-\sigma_s-2\sigma_s^2)Y_c }{(1+\sigma_c)\kappa^2 Y_s}
\left[
\begin{array}{ccc}
\frac{k_x^2 +\sigma_c k_y^2}{1-\sigma_c} & k_x k_y  & 0 \\
k_x k_y & \frac{\sigma_c k_x^2+k_y^2}{1-\sigma_c} & 0 \\
0&0 &  0
\end{array}
\right]
\end{equation}

\begin{eqnarray}
\label{Sijs}
\| \tilde S_{ij} \| = - \frac{\tilde w_s   (1-\sigma_s -2\sigma_s^2) }{\kappa^2 Y_s }
\left[
\begin{array}{ccc}
k_x^2 & k_x k_y \\
k_x k_y & k_y^2 \\
& &  \displaystyle \frac{-\sigma_c}{1-\sigma_c}
\end{array}
\right]\!\!,\;
\end{eqnarray}
where Young's modulus $Y_c$ and Poisson's ratio $\sigma_c$  of the coating are given by values within layer I.  The total dissipation
in this case will have the following structure,
\begin{equation}
\frac{W_{\rm diss}}{F_0^2} = l_1 \left[W_{11} \int w_1^2 d^2\vec x + 2W_{1s} \int w_1w_s d\vec x + W_{ss} \int w_s^2  d^2\vec x\right]\,,
\end{equation}
with the first term arising from dissipation in layer I that is due to strain and stress driven by $f_1$, the second term also arising from dissipation in layer I arising from cross terms between strains and stresses caused by $f_1$ and $f_s$, and the third term arises from dissipations throughout the entire coating, due to strain and stress caused by $f_s$. 
Here $W_{11}$ is the same as defined by Eq.~\eqref{wjj}, and
\begin{subequations}
\begin{align}
W_{js} &= \frac{(1-\sigma_s -2\sigma_s^2)(1-\sigma_j-2\sigma_j^2)}{2(1-\sigma_j)^2 Y_s} (\phi_B^j-\phi_S^j) \\
W_{ss}^{(j)} &=\frac{(1-\sigma_s-2\sigma_s^2)^2 Y_j }{(1-\sigma_j)^2 Y_s^2}\left[\frac{1-2\sigma_j}{2}\phi_B^j+\frac{1-\sigma_j+\sigma_j^2}{1+\sigma_j}\phi_S^j\right]
\end{align}
\end{subequations}
Note that we have added a superscript $(j)$ for $W_{ss}$ to indicate that here the dissipation is due to force applied on one thin layer alone. 

Here again, the dependences on $w_{1}^2$ and $w_s^2$ indicates that fluctuations at different transverse locations, $\vec x\neq \vec x'$, are uncorrelated, while  the $l_1$ in front of $W_{11}$, and the arbitrariness of $l_1 $ means that $S_{zz}$ fluctuations at different $z$ locations within the thin layers are uncorrelated.   The $l_1$ in front of both $W_{1s}$ and $W_{ss}$ indicates that all $S_{zz}$ within layer I are correlated with $z_s$ the same way, even though all of them are mutually uncorrelated. 

This allows us to extract the following
\begin{subequations}
\begin{align}
\label{sssx}
S_{z_sz_s}(\vec x,\vec x')
&=\frac{4k_BT}{3\pi f} \delta^{(2)}(\vec x -\vec x') \sum_j l_j W_{ss}^{(j)}\\
\label{sslx}
S_{S_{zz}\,s_z}(\vec x;\vec x',z')
&=\frac{4k_BT}{3\pi f}\delta^2(\vec x -\vec x') W_{js}\,.
\end{align}
Here for Eq.~\eqref{sslx}, $j$ is the layer with which $z'$ is associated; and this labeling is to help identify which material parameter to use in $W_{js}$. 

\end{subequations}

%
%
%

\subsection{The natomy of coating thermal noise}
\label{subsec:anatomy}

Here let us assemble  Eqs.~\eqref{sllx}, \eqref{sssx} and \eqref{sslx} from the previous sections, and write: 
\begin{widetext}
\begin{subequations}
\begin{align}
\label{sll}
S^{ij}_{S_{zz}  S_{zz}}(\vec{x},z;\vec{x}',z')
&=\frac{4k_BT}{3\pi f}\frac{(1+\sigma_j)(1-2\sigma_j)}{Y_j(1-\sigma_j)^2}
\Bigl{[}\frac{1+\sigma_j}{2}\phi_{Bj}+(1-2\sigma_j)\phi_{S j}\Bigr{]} \delta_{ij}\delta^{(2)}(\vec{x} -\vec{ x}')\delta(z-z') \\
\label{sss}
S_{z_sz_s}(\vec x,\vec x')
&=\frac{4k_BT}{3\pi f}\frac{(1-\sigma_s-2\sigma_s^2)^2}{Y_s^2}\sum_j \frac{Y_j l_j}{(1-\sigma_j)^2}\Bigl{[}\frac{1-2\sigma_j}{2}\phi_{B j}+\frac{1-\sigma_j+\sigma_j^2}{1+\sigma_i}\phi_{S j}\Bigr{]}\delta^{(2)}(\vec x -\vec x')\\
\label{ssl}
S_{z_s\,S_{zz}}(\vec x;\vec x',z')
&=\frac{2k_BT}{3\pi f}\frac{(1-\sigma_s-2\sigma_s^2)(1-\sigma_j-2\sigma_i^2)}{Y_s(1-\sigma_j)^2}[\phi_{B j}-\phi_{S j}]\delta^2(\vec x -\vec x')
\end{align}
\end{subequations}
\end{widetext}
Here we have assumed that $z$ belongs to the $i$-th layer and that $z'$ belongs to the $j$-th layer, respectively.  The thickness fluctuation of different layers are mutually independent [note the Kronecker delta in Eq.~\eqref{sll}], while thickness fluctuation of each layer is correlated with the height fluctuation of the coating-substrate interface [Eq.~(\ref{ssl})].  

Fluctuations described by Eqs.~\eqref{sll}--\eqref{sss} can be seen as driven by a set of microscopic fluctuations throughout the coating.  Suppose we have $3N$ thermal noise fields (i.e., 3 for each coating layer), $n_j^B(\mathbf{x})$, $n_j^{S_A}(\mathbf{x}) $ and $n_j^{S_B}(\mathbf{x}) $, all independent from each other, with
\begin{subequations}
\begin{align}
\label{SBB}
S_{n_j^B n_k^B} =&\frac{4k_B T (1-\sigma_j-2\sigma_j^2)}{3\pi f Y_j(1-\sigma_j)^2}\phi_B^j\delta_{jk}\delta^{(3)}(\mathbf{x}-\mathbf{x}'), \quad\;\\
S_{n_j^{S_A}n_k^{S_A}} =S_{n_j^{S_B} n_k^{S_B}}=&
\label{SSASA}
\frac{4k_B T (1-\sigma_j-2\sigma_j^2)}{3\pi f Y_j(1-\sigma_j)^2}\phi_S^j
\delta_{jk}\delta^{(3)}(\mathbf{x}-\mathbf{x}'),\quad \; 
\end{align}
\end{subequations}
and all other cross spectra vanishing. Here $j$ labels coating layer, the superscript $B$ indicates bulk fluctuation, while $S_A$ and $S_B$ label two types of shear fluctuations. The normalization of these fields are chosen such that each of these fields, when integrated over a length $l_j$ along $z$, have a noise spectrum that is roughly the same magnitude as a single-layer thermal noise.

Noise fields $n_j^B(\mathbf{x})$, $n_j^{S_A}(\mathbf{x})$ and $n_j^{S_B}$ can  be used to generate thickness fluctuations of the layers and the interface fluctuation \eqref{sll}--\eqref{sss} if we define\begin{equation}
\label{uzzeff}
u_{zz}(\vec x,z) = C^{ B}_j n_j^B(\vec x,z) +C^{ S_A}_j n_j^{S_A}(\vec x,z)
\end{equation}
and 
\begin{eqnarray}
\label{zeff}
z_s(\vec x) &= &\sum_j \int_{L_{j+1}}^{L_j} dz \bigg[  D^{ B}_j n_j^B(\vec x,z)  +D^{S_A}_j n_j^{S_A}(\vec x,z)  \nonumber\\
&&\qquad\qquad\quad+D^{S_B}_j n_j^{S_B}(\vec x,z) \bigg] 
\end{eqnarray}
For each coating layer, $C_j^B$ and $D_j^B$ are transfer functions from the bulk noise field  $n^B_j$ to its own thickness $\delta l_j$ and to surface height $z_s$, respectively; $C^{S_A}_j$ and $D^{S_A}_j$ are transfer functions from the first type of shear noise to thickness and surface height; finally $D^{S_B}_j$ is the transfer function from the second type of shear noise to surface height (note that this noise field does not affect layer thickness). Explicit forms of these transfer functions are listed in Table.~\ref{tab:transfer}.

\begin{figure}
\includegraphics[width=0.35\textwidth]{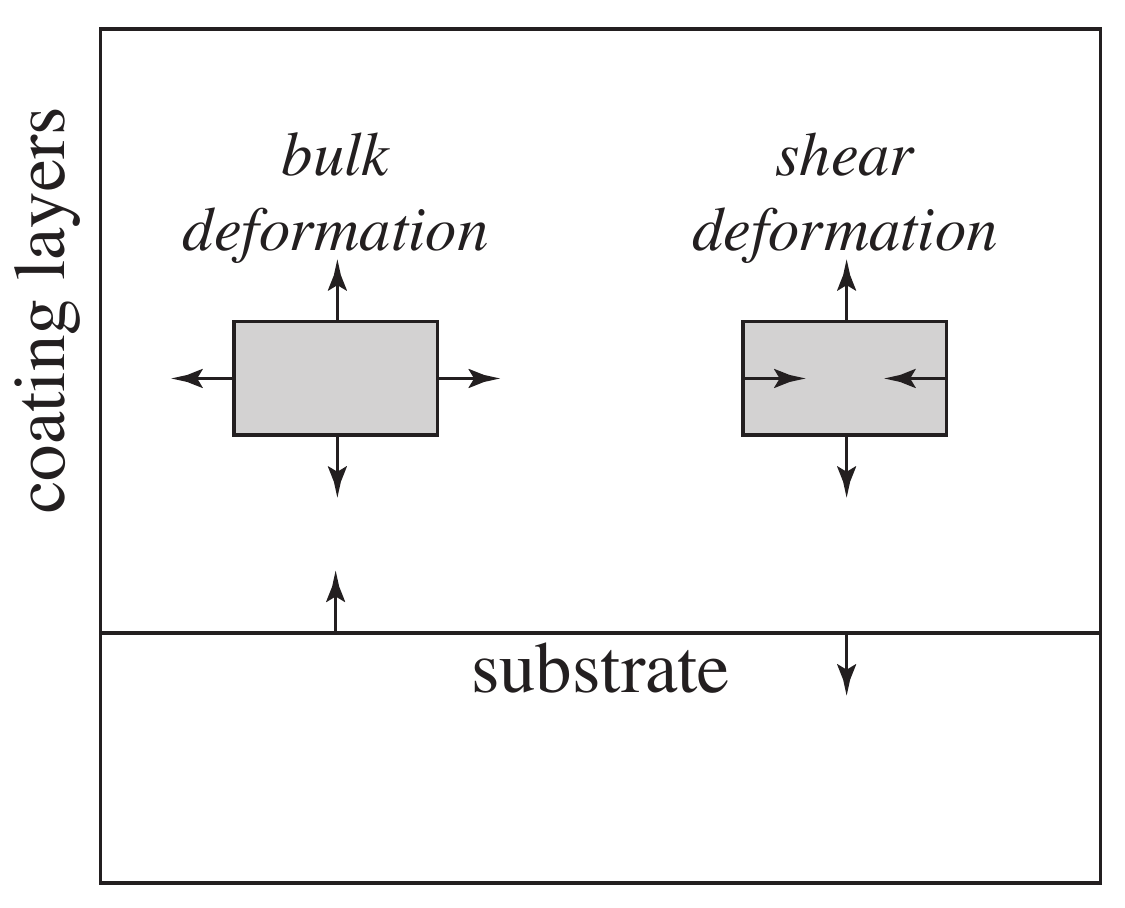}
\caption{\label{fig:correlation} \R{Illustration of the correlations between coating thickness $\delta l_j$ and the height of the coating-substrate interface, $z_s$.
On the left, for a bulk deformation: when a coating element is expanding, its expansion along the $x$-$y$ plane lifts the coating-substrate interface upwards, causing additional motion of the coating-air interface  correlated to that caused by the increase in coating thickness. On the right, a a particular shear mode:  when a coating element is expanding, its contraction along the $x$-$y$ plan pushes the coating-substrate interface downwards, causing addition motion of the coating-air interface anti-correlated to that caused by the increase in coating thickenss.}}
\end{figure}

\begin{table}
\begin{tabular}{ccc}
\hline\hline
& Thickness ($\delta_j$)  & Surface height ($z_s$)  \\ 
\hline
Bulk &  $\displaystyle C^{B}_j =\sqrt{\frac{1+\sigma_j}{2}}$ & 
$\displaystyle  D^B_j= \frac{1-\sigma_s-2\sigma_s^2}{\sqrt{2(1+\sigma_j)}} 
 \frac{Y_j}{Y_s}  $ \\
\begin{tabular}{c} 
Shear \\ A\end{tabular}  
& 
$\displaystyle  C^{S_A}_j = \sqrt{1-2\sigma_j}$ & 
$\displaystyle  D^{S_A}_j = -\frac{1-\sigma_s-2\sigma_s^2}{2\sqrt{1-2\sigma_j}}\frac{Y_j}{Y_s}$
\\
\begin{tabular}{c} 
Shear \\ B\end{tabular}  
 & (none) &  
$\displaystyle  D^{S_B}_j = \frac{\sqrt{3} (1-\sigma_j)(1-\sigma_s-2\sigma_s^2)}{2\sqrt{1-2\sigma_j}(1+\sigma_j)} \frac{Y_j}{Y_s}$ \\
\hline
\end{tabular}
\caption{Transfer functions from bulk and shear noise fields to layer thickness and surface height. \label{tab:transfer}}
\end{table}

Equations~\eqref{uzzeff} and \eqref{zeff} owe their simple forms to the underlying physics of thermal fluctuations:   

For {\it bulk noise}, i.e., terms involving $n_j^B$, the form of Eqs.~\eqref{uzzeff} and \eqref{zeff} indicates that the interface fluctuation  due to bulk dissipation is simply a sum of pieces that are directly proportional to the bulk-induced thickness fluctuations of each layer. This means the thermal bulk stress in a layer drive simultaneously the thickness fluctuation of that layer and a fluctuation of the coating-substrate interface.   The fact that $D_j^{B}$ and $C_j^{B}$ having the same sign means that when thickness increases, the interface also rises (with intuitive explanation shown in Figure~\ref{fig:correlation}).  This sign of correlation is generally unfavorable because the two noises add constructively towards the rise of the coating-air interface.  

For {\it shear noise}, the situation is a little more complicated, because unlike bulk deformations, there are a total of 5 possible shear modes. From Eq.~\eqref{eq:T1} and \eqref{eq:S1}, it is clear that $f_1$, applied on opposites of Layer I (Figure~\ref{force}), only drives the $xx+yy-2zz$ shear mode and the $xx+yy+zz$ bulk mode, while from Eq.~\eqref{Tijs} and \eqref{Sijs}, the force distribution $f_s$ drives three shear modes of  $xx-yy$, $xy+yx$, and $xx+yy-2zz$. This means while thermal shear stresses in the $xx+yy-2zz$ mode drives layer thickness and interface fluctuation simultaneously, there are additional modes of shear stress, $xx-yy$ and $xy+yx$, that only drives the interface without driving layer thickness.  Our mode $S_A$, which drives both layer thickness and interface height, therefore corresponds to the physical shear mode of $xx+yy-2zz$; our mode $S_B$, which only drives interface height, corresponds to the joint effect of the physical shear modes $xx-yy$ and $xy+yx$.  It is interesting to note that for $S_A$, its contributions to $\delta l_j$ and $z_s$ are anti correlated, because $C^{S_A}$ and $D^{S_A}$ have opposite signs. This is intuitively explained in Fig.~\ref{fig:correlation}.

As an example application of Eqs.~\eqref{uzzeff} and \eqref{zeff}, if we ignore  light penetration into the coating layers, namely, when thermal noise is equal to 
\begin{equation}
\xi^{\rm np} \equiv -z_s- \sum_j \delta l_j
\end{equation}
we have
\begin{eqnarray}
\xi^{\rm np} =-
\sum_j &&\!\!\! \int\limits_{L_j}^{L_{j+1}}dz\Big[\left(C_j^B+D_j^{B}\right) n_j^B \nonumber\\
&&\qquad\;+ \left(C_j^{S_A}+D_j^{S_A}\right) n_j^{S_A}
\nonumber\\
&&\qquad\;+D_j^{S_B} n_j^{S_B}  \Big]
\end{eqnarray}
in which contributions from each layer has been divided into three mutually uncorrelated groups, each arising from a different type of fluctuations.  Here we see explicitly that $C^B$ and $D^B$ sharing the same sign increases contributions from $n^B$, $C^{S_A}$ and $D^{S_A}$ having opposite signs suppresses contributions from $n^{S_A}$.

Finally, we note that in the spectral density of $\xi^{\rm np}$,  contributions directly from coating thickness will be proportional to $|C^B_j|^2$ and $|C^{S_A}_j|^2$, and hence proportional to $1/Y_c$, those from interface height will be $|D^B_j|^2$, $|D^{S_A}_j|^2$ and $|D^{S_B}_j|2$, and hence proportional to $Y_c/Y_s^2$, while those from correlations will be proportional to $C^B_j D^B_j$ and $C^{S_A}_j D^{S_A}_j$, and hence proportional to $1/Y_s$.  This confirms our anticipation at the end of Sec.~\ref{subsubsec:stn}.

%

\subsection{Full formula for thermal noise}
\label{subsec:fullformula}

As we consider light penetration into the coating, we resort to Eq.~\eqref{eq:xi}, and write:
\begin{eqnarray}
\label{thermal_simple}
&&\xi(\vec x)- i\zeta(\vec x) \nonumber\\
\!\!\!&=&\!\!\!-\sum_j \int_{z_{j+1}}^{z_{j}}dz\bigg\{
\left[
\left[
1+\frac{i\epsilon_j(z)}{2}
\right] C_j^B +D_j^B\right]  n_j^B (\vec x,z)
\nonumber\\ 
&&\qquad\qquad\quad \;\;
+\left[
\left[
1+\frac{i\epsilon_j(z)}{2}
\right] C_j^{S_A} +D_j^{S_A}\right]  n_j^{S_A} (\vec x,z)
\nonumber\\
&&\qquad\qquad\quad \;\;+D_j^{S_B} n_j^{S_B}(\vec x,z)  \Big]\bigg\}
\end{eqnarray}
Here spectra of independent fields $n_{j}^{B}$, $n_{j}^{S_{A}}$ and  $n_{j}^{S_{B}}$ have been given in Eqs.~\eqref{SBB}--\eqref{SSASA},  $\epsilon$ is defined in Eq.~\eqref{epsilon}, while the transfer functions $C$'s and $D$'s are listed in Table.~\ref{tab:transfer}.

We can then obtain the spectrum of phase noise (after averaging over the mirror surface, weighted by the power profile of the optical mode) as
\begin{eqnarray}
\label{Sxi}
S_{\bar\xi} &=& \sum_{j}\int_{z_{j+1}}^{z_{j}} \frac{dz}{\lambda_j} \left[\left[1-\mathrm{Im}\frac{\epsilon_{j}(z)}{2}\right] C_{j}^{B}+D_{j}^{B}\right]^{2} {S^{B}_j} \nonumber\\
&+& \sum_{j}\int_{z_{j+1}}^{z_{j}} \frac{dz}{\lambda_j} \left[\left[1-\mathrm{Im}\frac{\epsilon_{j}(z)}{2}\right] C_{j}^{S_A}+D_{j}^{S_A}\right]^{2} S^{S}_j \nonumber\\
&+& \sum_{j}\left[D_j^{S_B}\right]^2 \frac{l_j}{\lambda_j} S_j^{S} \nonumber\\
&\equiv& \sum_j q_j^B S_j^B + q_j^{S} S_j^{S}
\end{eqnarray}
 and spectrum of amplitude noise as
\begin{eqnarray}
\label{Szeta}
S_{\bar\zeta} &=& \sum_{j}\int_{z_{j+1}}^{z_{j}} \frac{dz}{\lambda_j} 
\,\bigg\{\left[C_{j}^{B}\mathrm{Re}\frac{\epsilon_{j}(z)}{2} \right]^{2} {S^{B}_j} 
\nonumber\\
&&\qquad\qquad\quad+ \left[C_{j}^{S_A}  \mathrm{Re}\frac{ \epsilon_{j}(z)}{2}\right]^{2} S^{S}_j\bigg\}
\end{eqnarray}
Here $\lambda_j$ is the wavelength of light in Layer $j$, and  we have defined
\begin{equation}
S_j^X \equiv \frac{4k_B T  \lambda_j \phi_X^j (1-\sigma_j-2\sigma_j^2)}{3\pi f Y_j(1-\sigma_j)^2 \mathcal{A}_{\rm eff}}\,,\quad X=B,S\,.
\end{equation}
which is at the level of coating thickness fluctuation of a  single layer of dielectrics with material parameters identical to layer $j$ and length equal to $\lambda_j$.  Note that the quantity $S_j^X$ only depends on the material properties (and temperature) of the layer, and is independent from length of that layer; the quantities $q_j^X$, on the other hand, give us the relative thermal-noise contribution of each layer in a dimensionless way.

 Note  that the reason for keeping the integrals in Eqs.~\eqref{Sxi} and \eqref{Szeta} is because $\epsilon$ has a $z$ dependence, which originates from the fact that the back-scattering contributions to $\delta\phi_j$'s and $\delta r_j$'s  a weighted integral of $u_{zz}$ within each layer [Cf.~\eqref{dphij} and \eqref{drj}]. 
%

\section{Effect of Light penetration into the coating}
\label{sec:penetration}

In this section, we synthesize results from Sec.~\ref{sec:components} and Sec.~\ref{sec:spectra}, and compute the full Brownian thermal noise for coating configurations.  We will illustrate how the light penetration affects the total noise in highly reflective coatings.

\subsection{Optics of multi-layer coatings}

For completeness of the paper, we briefly review how light penetration coefficient $\partial \log\rho/\partial \phi_j$ can be calculated. 
\begin{figure}[t]
\centering
\includegraphics[width=0.425\textwidth]{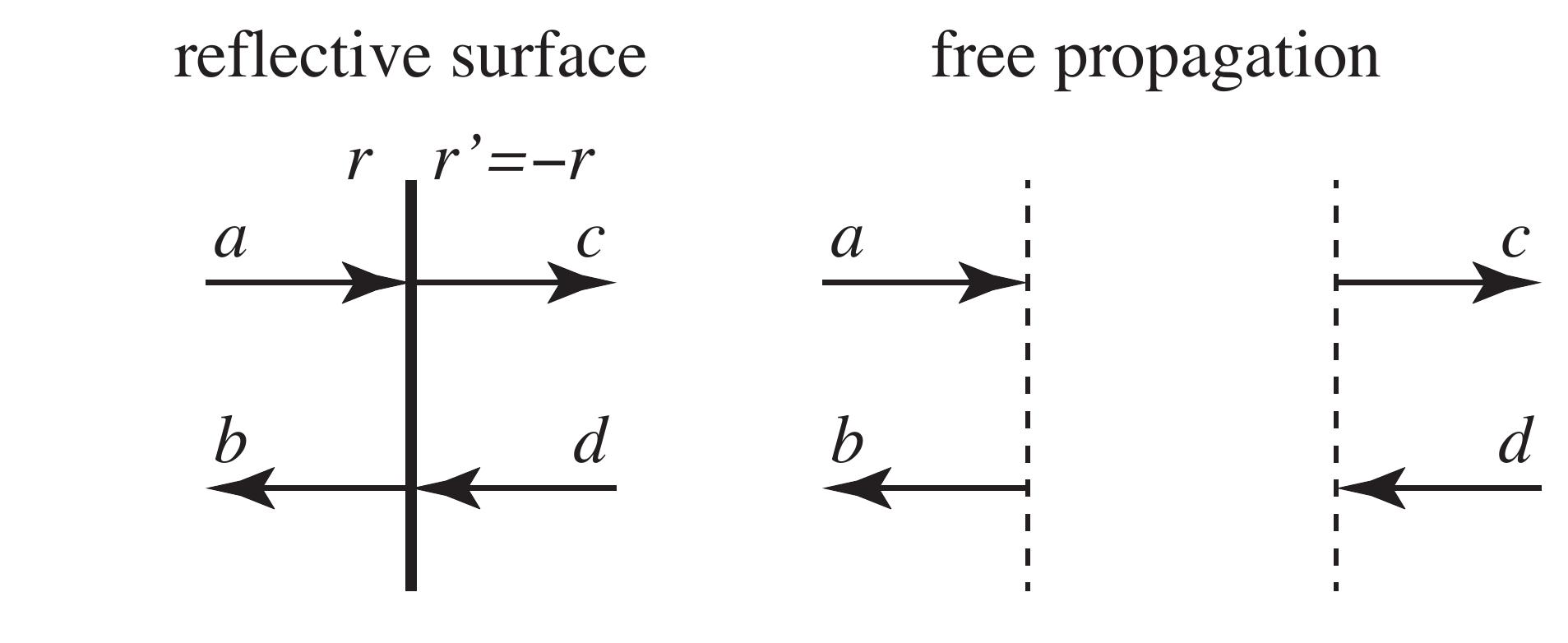}
\caption{\label{fig:matrix} Two basic transformations involved in solving for optical fields in a multi-layer coating.}
\end{figure}

\begin{table}
\begin{tabular}{l*{7}{c}r}
\toprule
  Parameter & & Tantala($\rm Ti_2O_5$)  & & Silica($\rm SiO_2$) \\
\hline
  Refractive index  && 2.07 \cite{para2}&&1.45 \cite{para2} \\
  Poisson's ratio && 0.23 \cite{para1}& & 0.17\cite{para1}\\
  Young's modulus (Pa) & &$1.4\times10^{11}$\cite{Martin} &&$7\times 10^{10}$\cite{para1}\\
  Loss angle ($\phi_B=\phi_S$) & & $2.3\times10^{-4}$\cite{harry2}& &$4.0\times 10^{-5}$\cite{penn} \\
  Photoelastic coefficient && -0.50 \cite{Nakagawa} && -0.41\cite{stone}\\
\hline
\end{tabular}
\caption{Baseline material parameters. \label{tab2}}
\end{table}

\begin{figure*}[t]
\includegraphics[width=\textwidth]{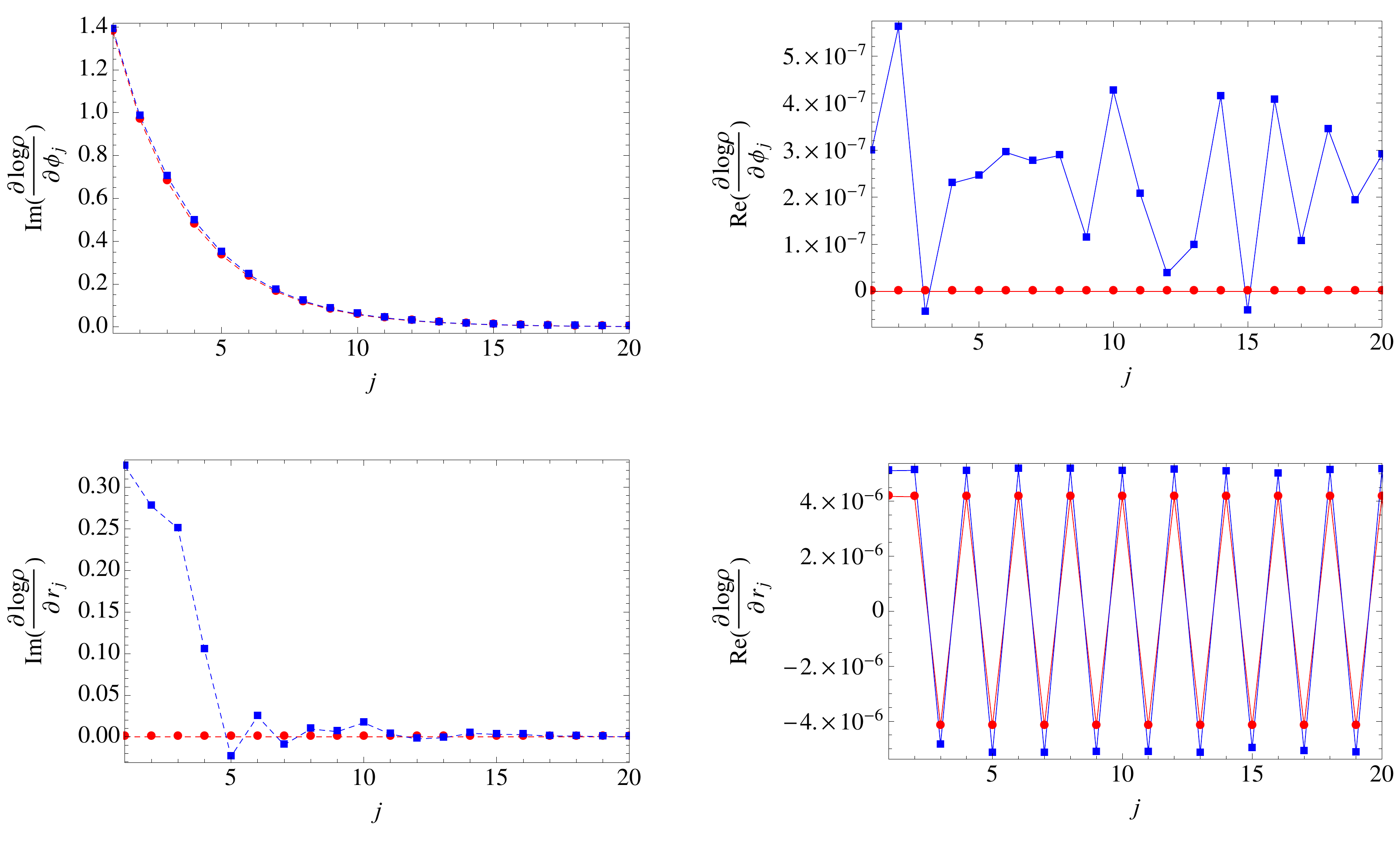}

\caption{Real (solid curves) and imaginary (dashed curves) parts of $\partial\log\rho/\partial\phi_j$ (upper panel) and $\partial\log\rho/\partial r_j$ (lower panel), for conventional (red curve) and Advanced LIGO (blue curve) coatings. [Note that $\mathrm{Re}(\partial\log\rho/\partial\phi_j)=0$ for conventional coating.] \label{fig:drho}}
\end{figure*}

\begin{figure*}[t]
\includegraphics[width=\textwidth]{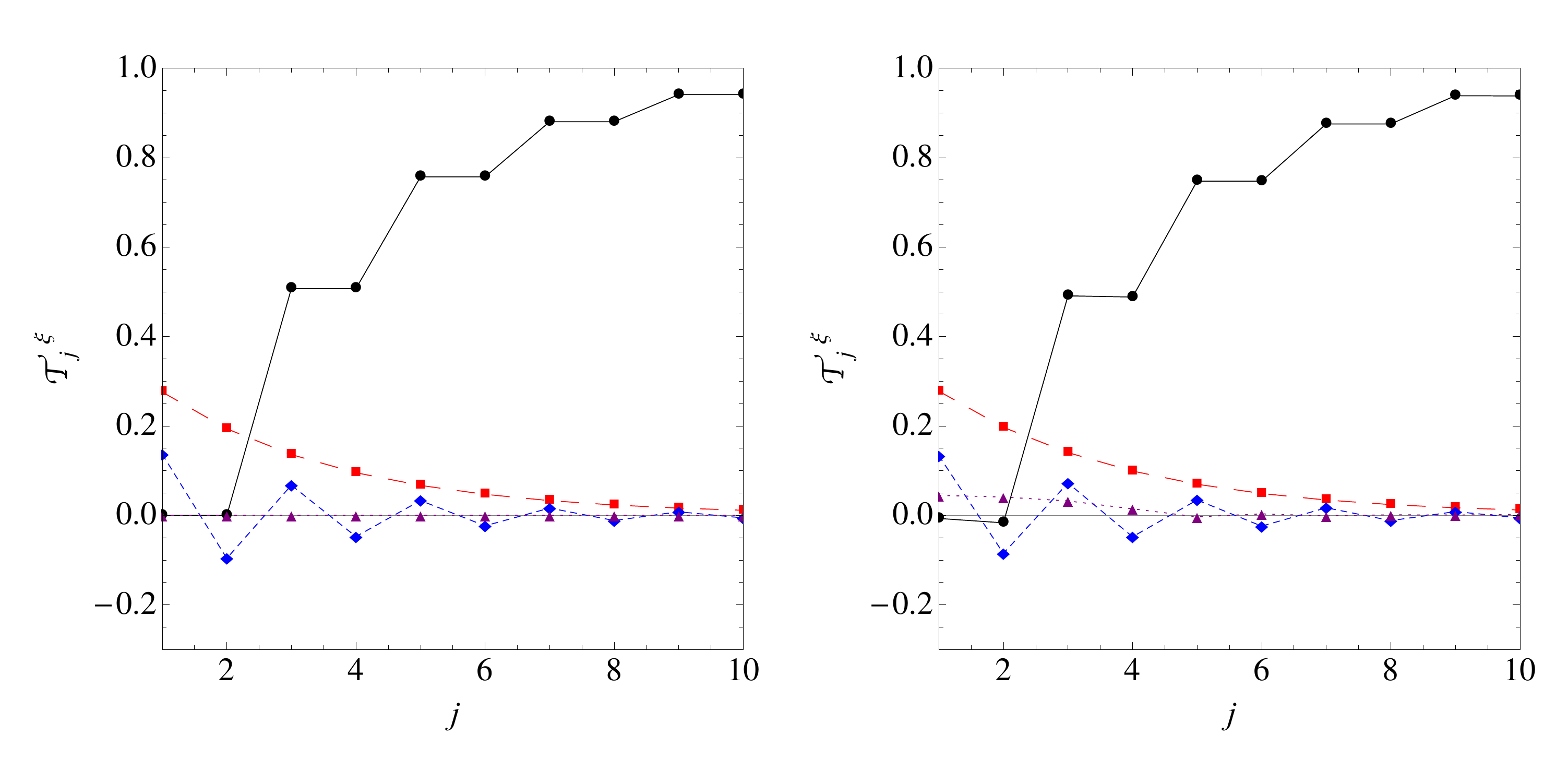}
\caption{Light penetration into the first 10 layers of a 38-layer coating (left panel for conventional coating and right panel for Advanced LIGO coating). We plot the non-photoelastic part of $\mathcal{T}_j$ in black sold curves, the photoelastic part of $\mathcal{T}_j^s$ in long-dashed red curves, as well as $\mathcal{T}_j^s$(scaled by rms value of $\delta l_j^c$ with respect to the rms value of $\delta l_j$, shown in short-dashed blue curves) and  $T_j^s$ (scaled by rms value of $\delta l^s_j$, shown in dotted purple curves).  \R{These plots indicate that for both structures, light penetration is restricted within the first 10 layers.} 
\label{fig:refplot}}
\end{figure*}

\begin{figure*}[t]
\includegraphics[width=\textwidth]{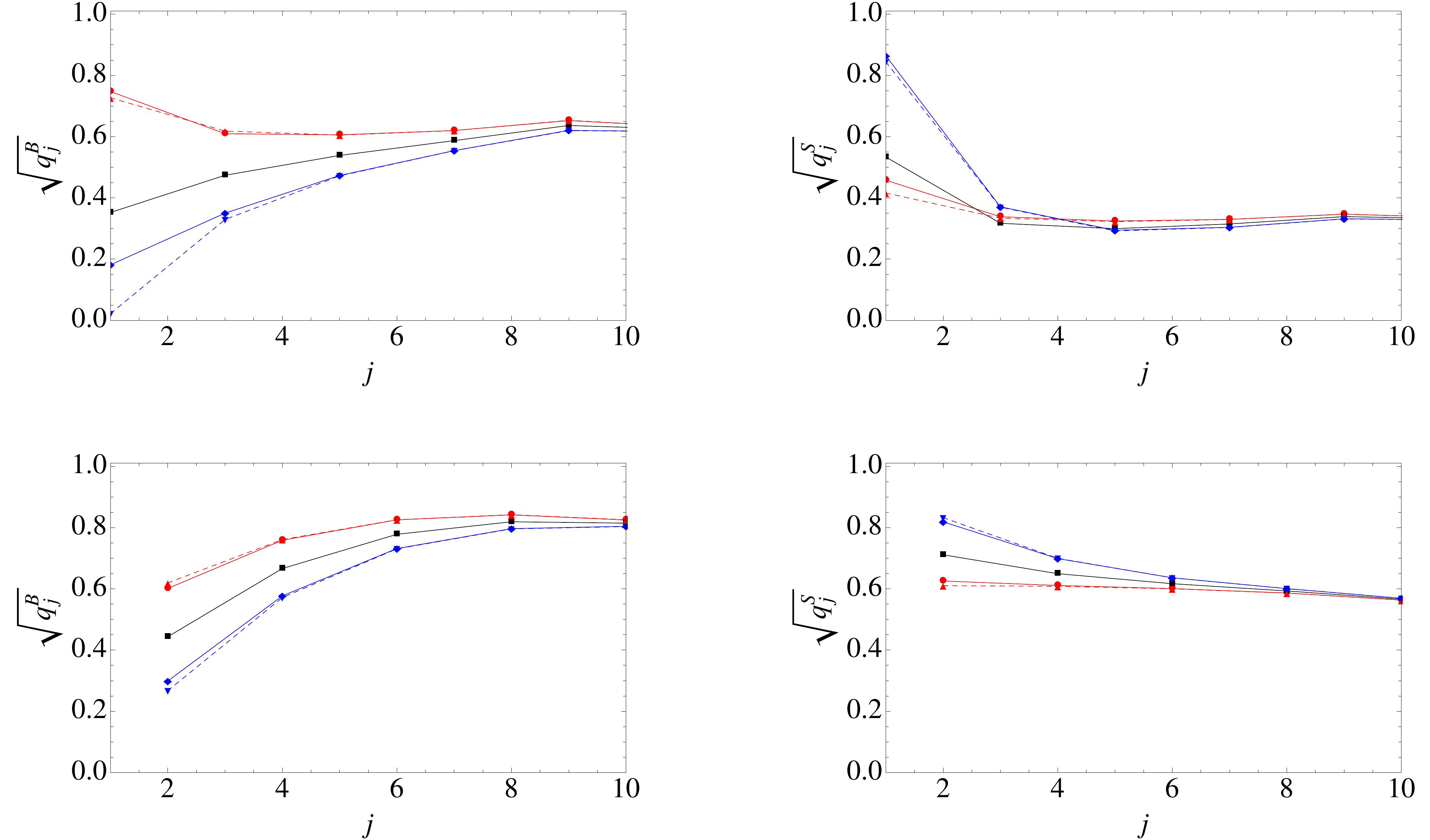}
\caption{A break-down of thermal noise contributions from silica (upper panels) and tantala (lower panels) layers, from bulk (left panels) and shear (right panels) losses.  Blue curves correspond to $\beta=-1$, black $\beta=0$ and red $\beta=1$. Dashed curves indicate results calculated without including \R{back-scattering effects}. 
\label{fig:thermal_layers}}
\end{figure*}

From an interface from layer $i$ to $j$ (here $j$ is either $i+1$ or $i-1$), we denote the reflectivity and transmissivity of different layers by $r_{ij}$ and $t_{ij}$: $r_{ij}^2+t_{ij}^2=1$. \be
r_{ij}=\frac{n_i-n_j}{n_i+n_j}
\ee
We also define $n_{N+1}=n_1$, since that is the refractive index of the substrate.

A matrix approach can be applied to solve for the amplitude of light inside the layers, when we view the coating as made up from two elementary transformations, each representable by a matrix. In this approach, instead of writing out-going fields in terms of in-going fields, one write fields to the right of an optical element in terms of those to the left. As illustrated in Figure~\ref{fig:matrix}, for reflection at an interface (left panel), we write
\begin{equation}
\left[\begin{array}{c}
c \\
d \\
\end{array}
\right]
\equiv \mathbf{R}_r
=\frac{1}{t}
\left[
\begin{array}{c c}
1 & -r \\
-r & 1 \\
\end{array}
\right]
\left[\begin{array}{c}
a \\
b \\
\end{array}
\right]
\end{equation}
On the other hand, for propagation across a gap with phase shift $\phi$, we have 
\begin{equation}
\left[\begin{array}{c}
c \\
d \\
\end{array}
\right]\equiv \mathbf{T}_\phi
=\left[
\begin{array}{c c}
e^{i\phi} &  \\
 & e^{-i\phi} \\
\end{array}
\right]
\left[\begin{array}{c}
a \\
b \\
\end{array}
\right]
\end{equation}

In this way, assuming the input and output field amplitude at the top surface of a multi-layer coating to be $v_1$ and $v_2$, and writing those right inside the substrate to be $s_1$ and $s_2$, we have
\begin{equation}
\left[\begin{array}{c}
s_1 \\
s_2 \\
\end{array}
\right]
=\left[
\begin{array}{c c}
M_{11} & M_{12} \\
 M_{21}& M_{21} \\
\end{array}
\right]
\left[\begin{array}{c}
v_1 \\
v_2 \\
\end{array}
\right]
={\mathbf{ M}}\left[\begin{array}{c}
v_1 \\
v_2 \\
\end{array}
\right]
\end{equation}
where $\mathbf{M}$ is given by
\begin{equation}
\mathbf{M} = \mathbf{R}_{r_{N,N+1}} \mathbf{T}_{\phi_{N-1}} 
\mathbf{R}_{r_{N-1,N}} \ldots\mathbf{R}_{r_{12}} \mathbf{T}_{\phi_1} \mathbf{R}_{r_{01}}
\end{equation}
The complex reflectivity is given by
\begin{equation}
\rho =-M_{21}/M_{22}
\end{equation}

\subsection{Levels of light penetration in Advanced LIGO ETM Coatings}
\label{subsec:penetration}

In Advanced LIGO, the coating stack is made from two alternating layers of materials: $\mathrm{Si}\mathrm{O}_2$ ($n_1=1.45$) and $\mathrm{Ta}_2\mathrm{O}_5$ ($n_2=2.07$). Here we consider the End Test-mass Mirror (ETM).  In order to achieve very high reflectivity, \R{the coating is made of 19 successive pairs of alternating $\mathrm{Si}\mathrm{O}_2$ and $\mathrm{Ta}_2\mathrm{O}_5$ layers, all $\lambda/4$ in thickness except the top one, which is $\lambda/2$.}    We will refer to this as the {\it conventional coating}.    An alternative design has been made to allow the coating to operate at both 1064\,nm and 532\,nm. We shall refer to this as the {\it Advanced LIGO coating} (see Appendix.~\ref{app:adv})~\cite{aLIGO:coatingdesign}. 

In Fig.~\ref{fig:drho}, we plot real and imaginary parts of $\partial \log\rho/\partial\phi_j$ and $\partial \log\rho/\partial r_j$, for both conventional and Advanced LIGO coating.  Here we note that the real parts of these derivatives are at the order of $10^{-6}$, which means $\bar\zeta$ \R{is less than} $\bar\xi$ by  6 orders of magnitude.  This, together with considerations in Sec.~\ref{subsec:amp}, will make amplitude coating noise negligible.

In Eq.~\eqref{xi:terms}, we have divided contributions to $\xi$ into four terms, the first, $z_s$, is the height of the coating-substrate interface, while the other three are related to fluctuations in layer thickness, $\delta l_j$, $\delta l_j^c$ and $\delta l_j^s$, see Eqs.~\eqref{xi:terms}--\eqref{Tjs}.  We can illustrate the effect of light penetration by showing the relative size of these three contributions for each layer.   In Figure~\ref{fig:refplot}, we carry out this illustration, for conventional coating on the left panel and for Advanced LIGO coating on the right. 
%
%
We use solid black line to indicate the non-photoelastic part of $\mathcal{T}^{\xi}_j$ [i.e., term not containing $\beta_j$, see Eq.~\eqref{Tjxi}], and we use red-long-dashed, blue-short-dashed, and purple-dotted curves to indicate the photoelastic part of $\mathcal{T}^\xi_j$,  $\mathcal{T}_j^{\xi c} \sqrt{\langle (\delta l_j^c)^2\rangle/\langle (\delta l_j)^2\rangle}$ and $\mathcal{T}_j^{\xi s }\sqrt{\langle (\delta l_j^s)^2\rangle/\langle (\delta l_j)^2\rangle}$, respectively.
The weighting factors,
\begin{eqnarray} 
\sqrt{\langle (\delta l_j^c)^2\rangle/\langle (\delta l_j)^2\rangle} = \frac{1}{\sqrt{2}}\sqrt{1+\frac{\sin4\phi_j}{4\phi_j}} \,,\\
\sqrt{\langle (\delta l_j^s)^2\rangle/\langle (\delta l_j)^2\rangle }= \frac{1}{\sqrt{2}}\sqrt{1-\frac{\sin4\phi_j}{4\phi_j}}  \,,
\end{eqnarray}
have been added for $\mathcal{T}_j^{\xi c}$ and  $\mathcal{T}_j^{\xi s}$, respectively, to correct for the fact that $\delta l_j^c$ and $\delta l_j^s$ have different r.m.s.\ values compared to $\delta l$.   Because of the lack of experimental data, we have assumed $\beta_j =-0.4$ identically.  Note that in order to focus on the effect of light penetration, we have only showed the first 10 layers. 

In the figure, the effect of light penetration into the coating layers is embodied in the deviation of the black solid curve from unity in the first few layers, and in the existence of the other curves.    Although we cannot perceive the correlation between these contributions, we can clearly appreciate that only the first few layers are penetrated, and that the total effect of light penetration will be small.  We should also expect the effect of photoelasticity (dashed curves) to be small, and the effect of back-scattering (which gives rise to $\mathcal{T}_j^{\xi c}$ and $\mathcal{T}_j^{\xi s}$, blue and purple dashed curves) even smaller.

\subsection{Thermal noise contributions from different layers}
Let us now examine the breakdown of the total coating noise by plotting the coefficients $q_j^B$ and $q_j^S$ in Eq.~\eqref{Sxi}.  In Fig.~\ref{fig:thermal_layers}, we plot silica contributions on top panels, and tantala contributions on lower panels, with bulk contributions on left panels, and shear contributions on right panels.  Here we use the baseline parameters shown in Table~\ref{tab2}. As it turns out, the results for conventional and Advanced LIGO coatings are hardly distinguishable from each other --- therefore we only use the Advanced LIGO coating. The red curve uses $\beta=-1$, black uses $\beta=0$ and blue uses $\beta =1$. Superimposed onto the solid lines are dashed lines of each type, calculated without introducing the back-scattering terms; the effect is noticeable for the first few layers.

\section{Dependence of thermal noise on material parameters }
\label{sec:dependence}
Experimental knowledge of coating materials is limited. \B{ Most notably, values of Young's moduli and Poisson's ratios of the coating materials are still uncertain}, while only {\it one combination} of the two loss angles have been experimentally measured by ring-down experiments.  \R{In this section, we explore the possible variation in coating Brownian noise, away from the baseline configuration (Table~\ref{tab2}),  considering these uncertainties.  We shall use the Advanced LIGO coating structure mentioned in the previous section.} 

\begin{figure}
  \includegraphics[width=2.75in]{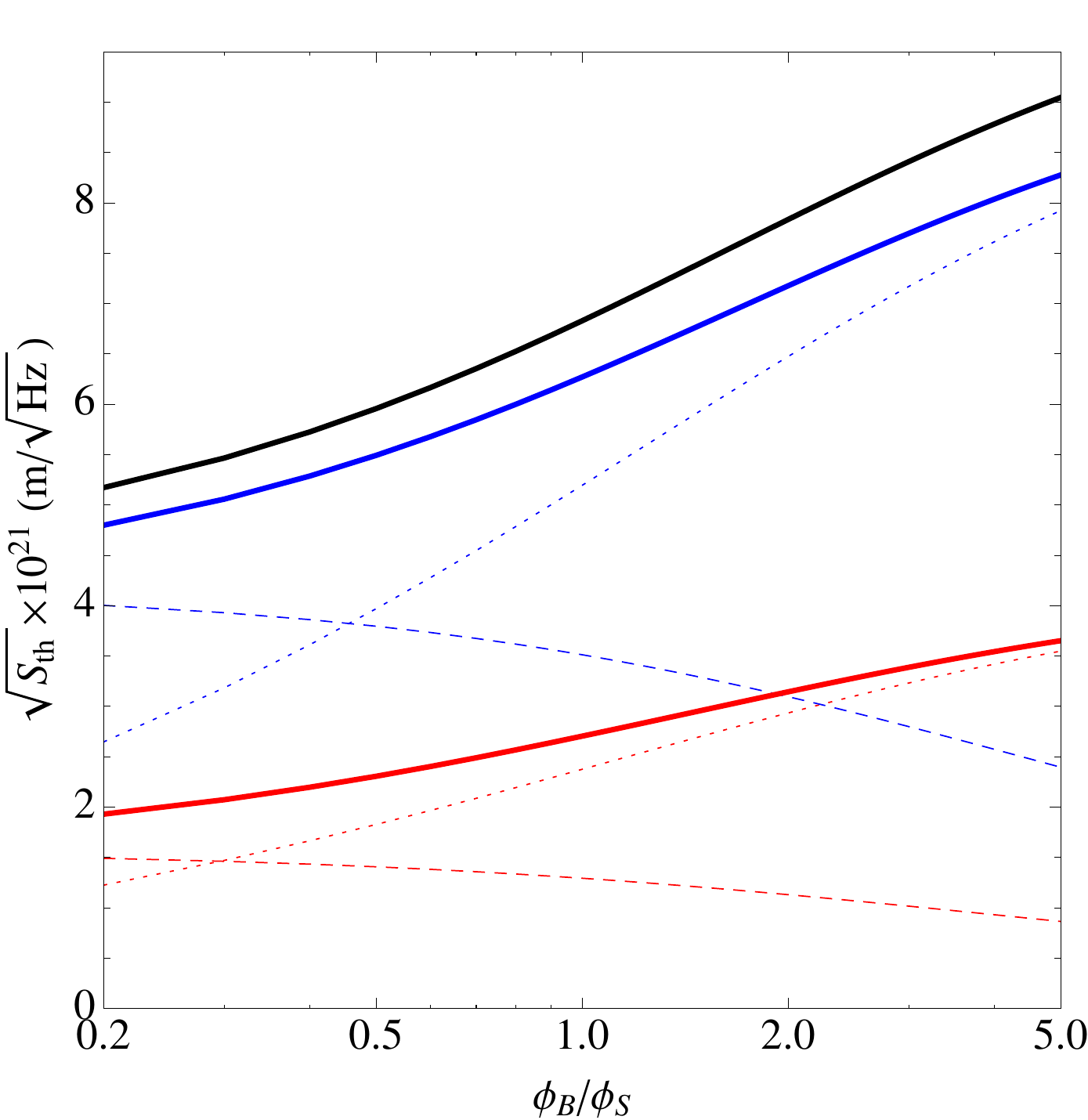}
       \caption{(Color Online) Variations in thermal noise contributions when $\phi_B/\phi_S$ is varied.  Contributions from tantala layers is shown in blue, those from silica layers are shown in red. The total thermal noise is in black. Bulk contributions are shown in dotted curves, while shear contributions are shown in dashed curves. \label{fig:baseline_zeta}}
\end{figure}

\begin{figure}
\includegraphics[width=2.75in]{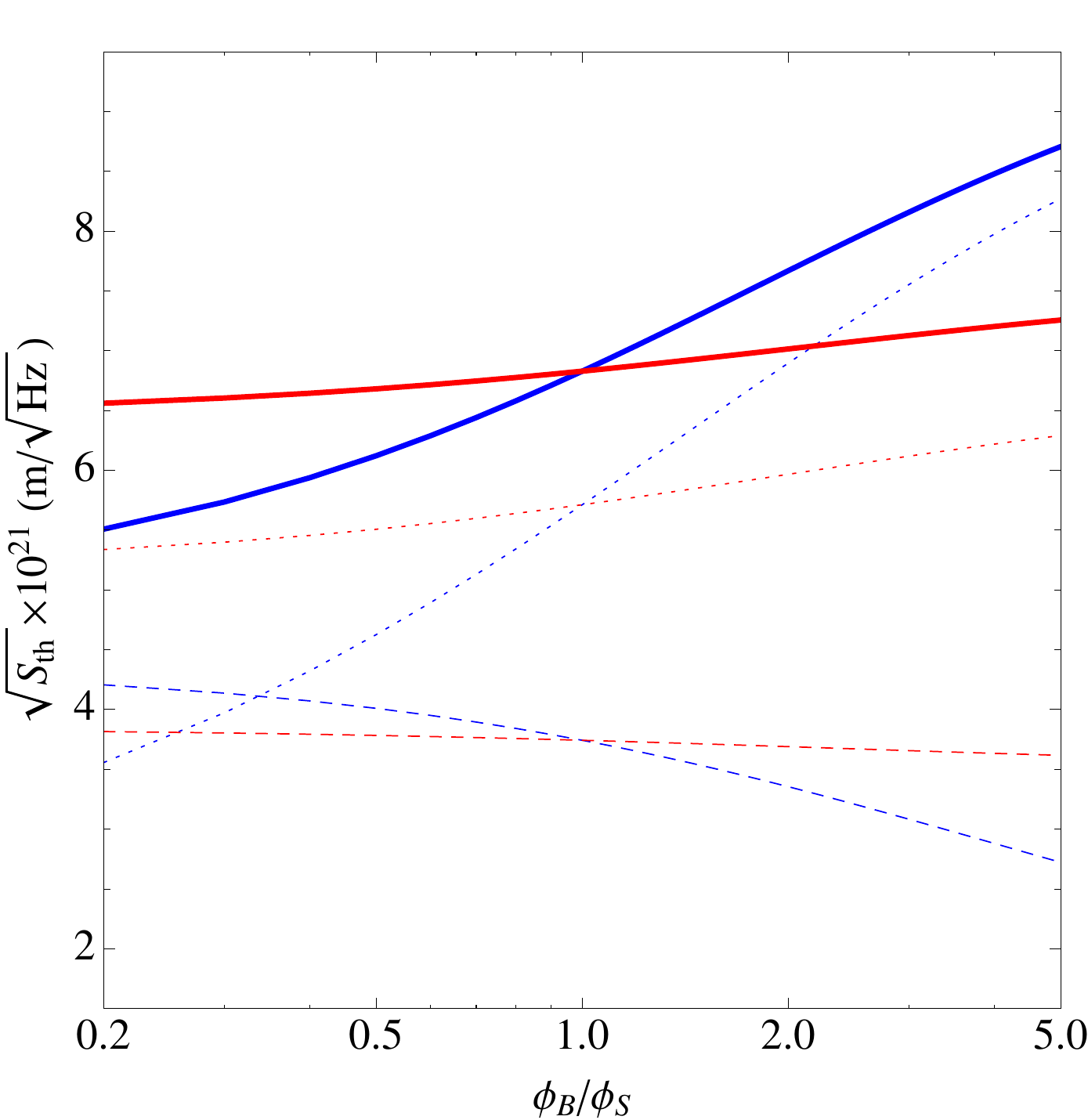}
\caption{(Color Online) Variations in total noise when $\phi_B/\phi_S$ is varied: (solid) total noise, (dotted) total bulk noise, (dashed) total shear noise. The red (blue) curve corresponds to only varying $\phi_B/\phi_S$ for tantala (silica). With $\phi_B/\phi_S$ of tantala or silica varying from 0.2 to 5, the change in total noise is $58.1\%$ and $10.6\%$
respectively. 
\label{fig:loss_total}}
\end{figure}

\subsection{Dependence on Ratios Between Loss Angles}
\label{subsec:lossangle}

In the baseline (Table~\ref{tab2}), we have assumed that $\phi_B$ and $\phi_S$ are equal, but this is only out of our ignorance: experiments have only been able to determine one particular combination of these two   angles.  We now explore the consequence of having these loss angles not equal, while keeping fixed the combination measured by ring down rate of drum modes [see Eq.~\eqref{totalloss}].

In Figure~\ref{fig:baseline_zeta}, while fixing all other baseline parameters, we plot how each type of thermal noise (i.e., silica vs tantala, bulk vs shear) varies when the ratio $\phi_B/\phi_S$ for both tantala and silica layers varies between 1/5 and 5.  We use blue for tantala, red for silica, dotted for bulk, dashed for shear, and solid for the total of bulk and shear.  In this configuration, tantala layers' contribution to thermal noise always  dominate over silica layers, mainly due to the higher loss angle.  As we vary the ratio between the loss angles, there is moderate variation of thermal noise.  For the dominant tantala, as $\phi_B/\phi_S$ vary from 1/5 to 5, there is a 30\% change in thermal noise, while for silica, the change is a more significant 68\%.  

\begin{figure}
       \includegraphics[width=2.75in]{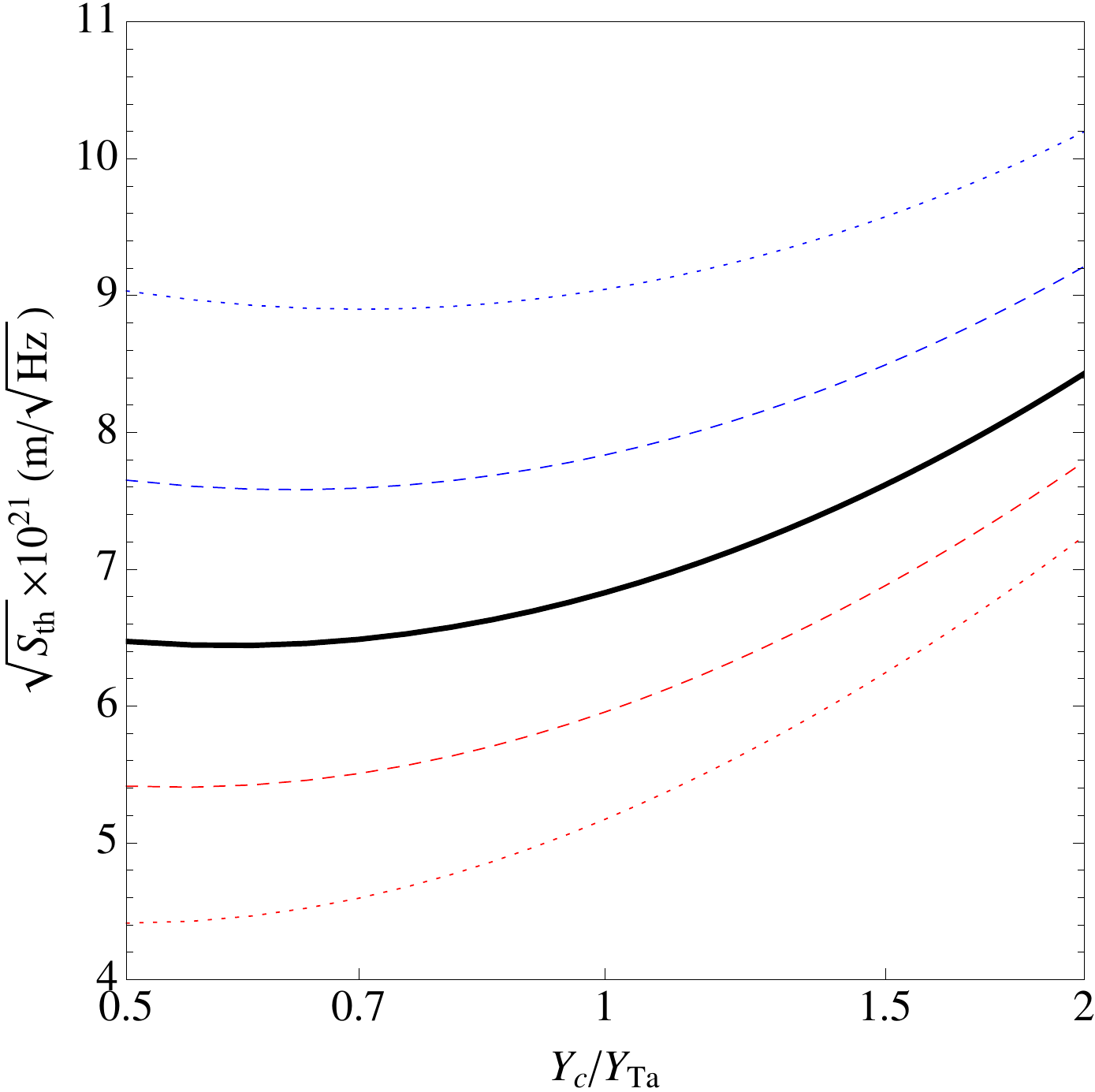}
               \caption{Thermal noise contribution from tantala, as its Young's modulus deviates from 
                              baseline value, for $\phi_B/\phi_S$=5 (blue dashed), 2 (blue dotted), 1 (black solid), 
                               1/2 (red dotted), and 1/5 (red dashed).}
   \label{fig:varyingYoungs}
\end{figure}

\begin{figure}
        \includegraphics[width=2.75in]{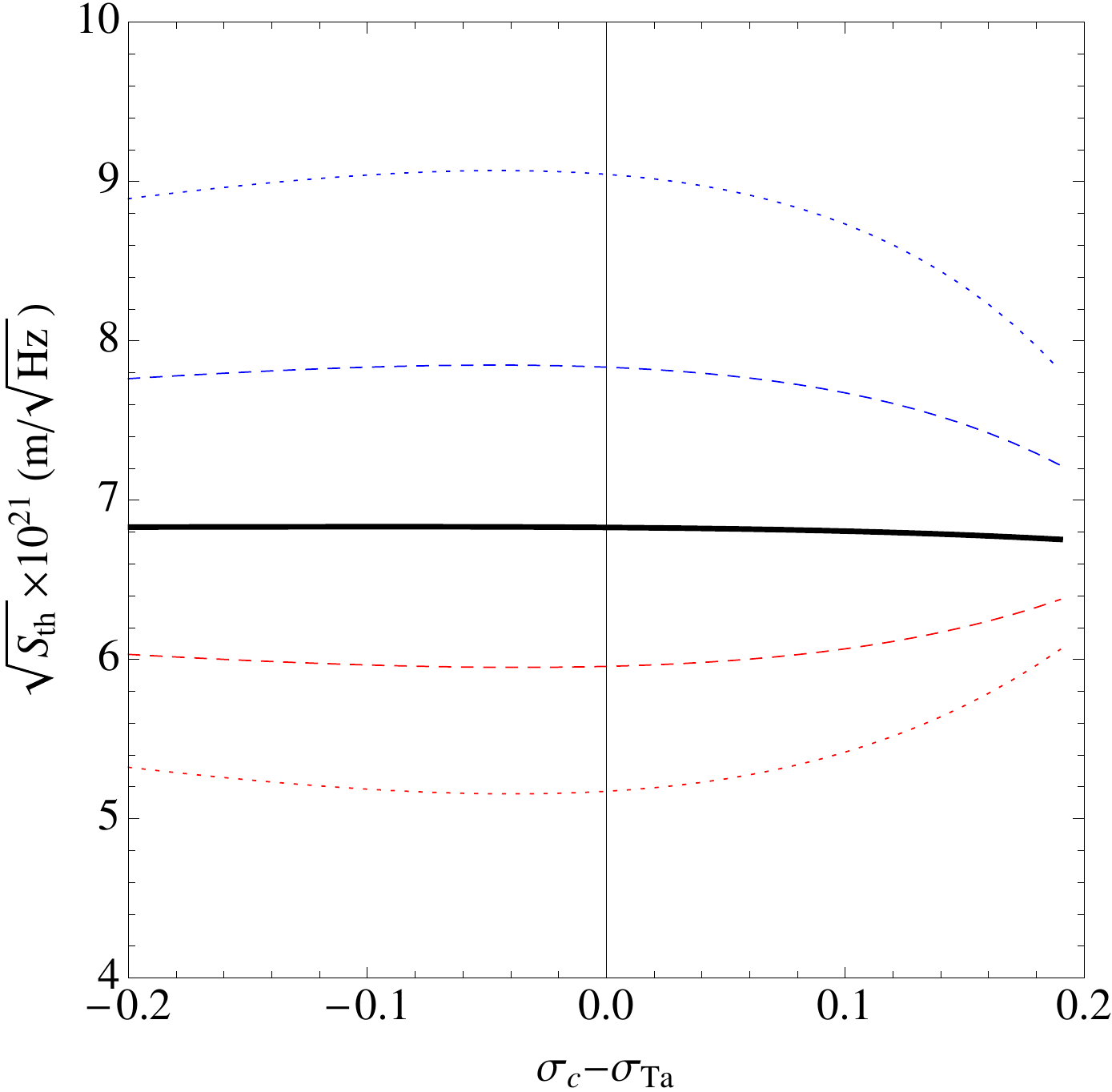}
                \caption{Thermal noise contribution from tantala, as its Poisson's ratio deviates from baseline value, 
                                for $\phi_B/\phi_S$=5 (blue dashed), 2 (blue dotted), 1 (black solid), 1/2 (red dotted), 
                                and 1/5 (red dashed).}
      \label{fig:varyingPoisson} 
\end{figure}

\begin{figure*}[t]
\includegraphics[width=\textwidth]{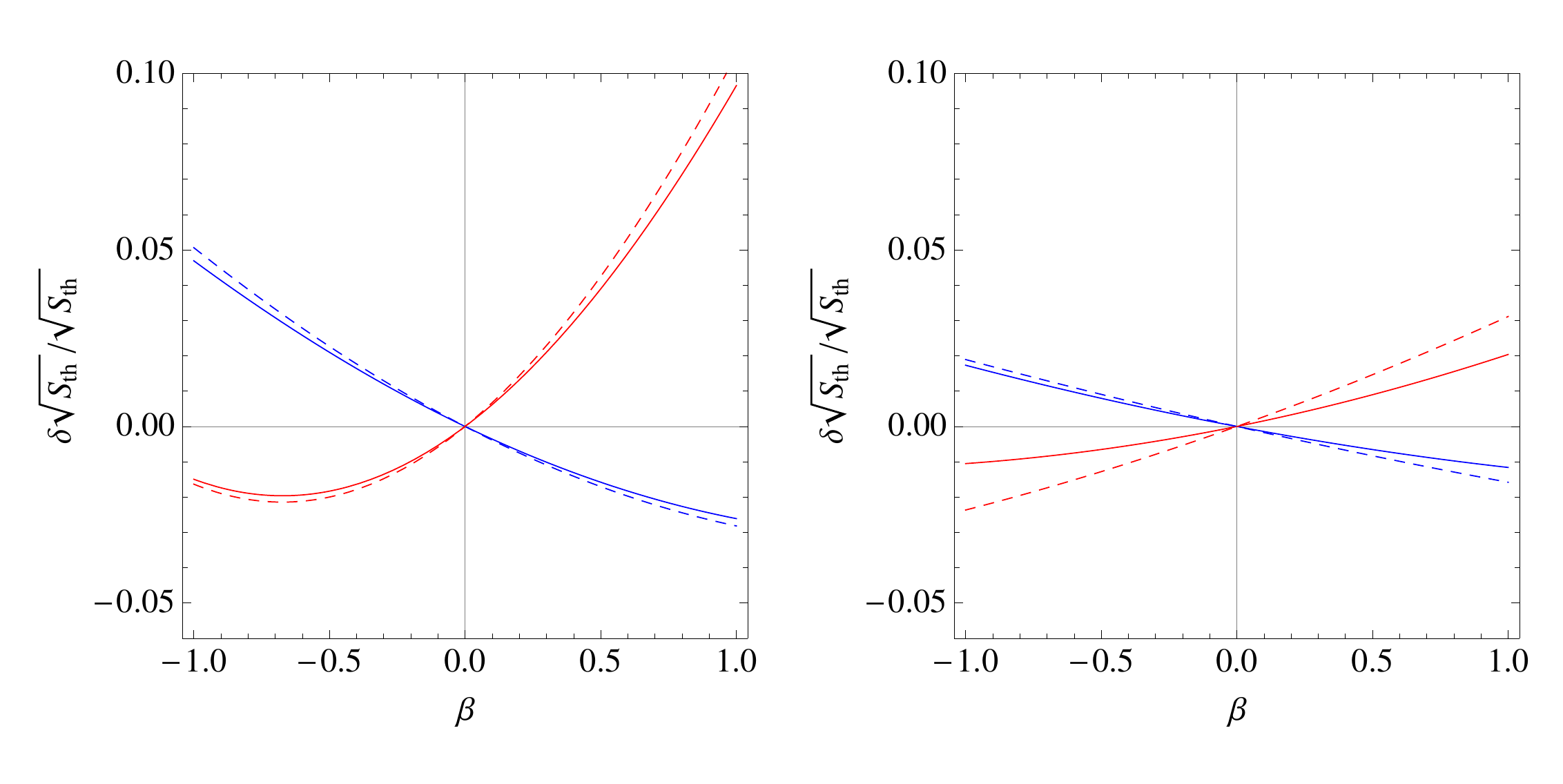}
\caption{Fractional change in the contribution to thermal noise from all silica layers  (left panel)  and all tantala layers (right panel), due to bulk (blue) and shear (red) loss.  Dashed lines indicate results calculated without including back-scattering terms. 
\label{fig:photoelastic}}
\end{figure*}

\R{As we see from Fig.~\ref{fig:baseline_zeta}, a larger value of $\phi_B/\phi_S$ gives rise to higher bulk, lower shear, and higher total noise --- this is reasonable because  bulk fluctuations drive correlated noise between layer's thickness and the height of coating-substrate interface, while shear fluctuations drive anti-correlated noise, as shown in Fig.~\ref{fig:correlation}.  }

\R{Moreover, the fact that} variation is more significant for silica layers can be explained when we recall that thickness-induced thermal noise is proportional to $1/Y_c$, while surface-height-induced thermal noise is proportional to $Y_c/Y_s^2$.  For silica layers, $Y_c$ is assumed to be equal to $Y_s$, so the two types of noise being added (bulk) or subtracted (shear) are more comparable in magnitude; by contrast, the Young's modulus of tantala layers is significantly higher than that of the substrate, causing the noise to be dominated by fluctuations of the height of the coating-substrate interface, therefore making correlations between the two types of noise less important. 

In Fig.~\ref{fig:loss_total}, we plot variations in the total noise as we vary $\phi_B/\phi_S$ for silica layers (blue) or tantala layers (red) only, and fix the other one. It shows that the 
variance of tantala's loss angle will generate larger change of the total noise.

\subsection{Dependence on Young's moduli and Poisson's ratios}
\label{subsec:depyoung}
\B{ Since the Young's modulus and Poisson's ratios of coating materials, especially of tantala, are also uncertain.}  In Fig.~\ref{fig:varyingYoungs}, we plot variations of tantala thermal noise when its Young's modulus varies from the baseline value by up to a factor of 2, for $\phi_B/\phi_S=$  0.2, 0.5, 1, 2 and 5.  The noise is seen to vary by $\sim$15\% as Young's modulus varies by a factor of $\sim 2$. 

We can also explain the way the thermal noise varies as a function of $Y_c$.  Starting from the baseline value, a lower $Y_c$ leads to a lower thermal noise, until $Y_c$ becomes comparable to $Y_s$ (which we fix at the baseline value, equal to $0.5 Y_{\rm Ta}$), and starts to increase again. Such a behavior is reasonable because thickness noise spectrum and interface noise spectrum are proportional to $\sim 1/Y_c$ and $\sim Y_c/Y_s^2$, respectively --- as we decrease $Y_c$ from the baseline $Y_{\rm Ta}$ value, we transition from interface fluctuation being dominant towards equal amount of both noises (which gives a minimum total noise), and then towards thickness fluctuation becoming dominant.

In Fig.~\ref{fig:varyingPoisson}, we explore the effect of varying coating Poisson's ratio, for the same values of $\phi_B/\phi_S$ chosen in Fig.~\ref{fig:varyingYoungs}.  In the baseline assumption of $\phi_B=\phi_S$, when bulk and shear have the same level of loss, thermal noise does not depend much on Poisson's ratio.  However, if $\phi_B/\phi_S$ turns out to differ significantly from $1$, and if Poisson's ratio can be larger than the baseline value by more than $\sim 0.1$, then thermal noise can vary by  $\sim 10\%$.

\subsection{Dependence on Photoelastic Coefficients}
\label{subsec:photoelastic}
\R{Photoelastic properties of the coating materials are not yet well known.}  In Fig.~\ref{fig:photoelastic}, we plot the fractional change in thermal noise, separately for silica (left panel) and tantala (right pane), and for bulk (blue) and shear (red) losses, when we vary $\beta$ between -1 and +1. Dashed curves are obtained ignoring back-scattering effects. 

It is interesting to note that for small values of $\beta$, the dependence of noise on $\beta$ have different trends for bulk and shear contributions.  This is also related to the different types of correlations between thickness and interface height fluctuations.  As we can see from the Figure, the effect of varying $\beta$ is small, since it only affects thermal noise due to light penetration into the first few layers.  If bulk and shear losses are indeed comparable, then cancelation between these two types of noises \R{(especially for the more lossy tantala layers) will likely make the photo elastic effect completely negligible.  Even in the case when one particular type of loss dominates shall we expect at most $\sim$2\% contribution from photo elasticity of the more lossy tantala --- if we further assume that $|\beta| \sim 1$ [right panel of Fig.~\ref{fig:photoelastic}]. }

\subsection{Optimization of Coating Structure}
\label{sec:optimization}
Although a standard highly reflective coating consists of $\lambda/4$ layers of alternating material capped by a $\lambda/2$ layer, this structure can be modified to lower thermal noise while still maintaining a high reflectivity for the 1064\,nm carrier light, e.g., as shown by Agresti et al.~\cite{pinto}.  As their results have indicated, for baseline coating parameters and neglecting light penetration into the coating layers~\cite{harry}, the optimal structure is more close to a stack of pairs of $\lambda/8$ ($\mathrm{Ta}_2\mathrm{O}_5$) and $3\lambda/8$ ($\mathrm{Si}\mathrm{O}_2$) layers, capped by a $\lambda/2$ ($\mathrm{SiO}_2$) layer.  This alternative coating structure shortens the \R{total thickness of the more lossy tantala layers, while  maintaining a high reflectivity for the light.}  The {\it Advanced LIGO type} coating given in Appendix~\ref{app:adv}, on the other hand, has been optimized considering reflectivity at both 1064\,nm and 532\,nm, as well as thermal noise --- although light penetration into the layers have not been considered. 

In this section, we carry out a numerical optimization taking penetration into account.  We first fix the number $N$ of layers ($N$ is even, so we have $N/2$ pairs), and then for $N$, we use the Lagrange multiplier method to search for the constrained minimum of $S_{\rm th}$, fixing $T_{1064}$ and $T_{532}$, namely the power transmissivity, $1-|\rho|^2$ assuming the coating is lossless, evaluated at 1064\,nm and 532\,nm, respectively. The quantity we seek to minimize (or, the {\it cost function}) is
\begin{equation}
y\equiv \sqrt{S_{\rm th}} + \mu_1 T_{1064}+ \mu_2 (T_{532}-5\%)^2
\end{equation}
As we vary $\mu_1$ and $\mu_2$ and minimizing $y$, we obtain the constrained minimum of $\sqrt{S_{\rm th}}$ for different pairs of $(T_{532},T_{1064})$.  The aim is to obtain a series of coating configurations with approximately $5\%$ transitivity for 532\,nm, and with minimized thermal noise for a variable 3 -- 20\,ppm transmissivity for 1064\,nm.  \R{(Note that the choice of the cost function contains a certain level of  arbitrariness.)}

Since we are going to carry out minimization for a large number of multipliers over a large number of degrees of freedom, we have chosen to proceed gradually allowing only the first $n$ pairs and last $n$ pairs of layers to vary, while maintaining the same pair structure for $N/2-n$ pairs in the middle (repeating doublets). In other words, our coating structure looks like:
\begin{equation*}
\underbrace{\quad\mbox{free}\quad}_{\mbox{$2n$ layers}}  \quad 
\underbrace{\mbox{repeating pair}}_{\mbox{$N-2n$ layers}} \quad \underbrace{\quad\mbox{free}\quad}_{\mbox{$2n$ layers}}  
\end{equation*}
In this work, we found that it suffices to choose $n=2$ (which corresponds to optimizing over 10 parameters); further increasing $n$ does not lead to noticeable improvements.   During our numerical optimization, we have adopted the {\it downhill simplex method}~\cite{Nelder:amoeba, NR:amoeba}. 

\begin{table*}[ht]
\begin{tabular}{|c|c|cccc|cc|cccc|c|c|c|c|c|}
\hline
target & &
\multicolumn{10}{c|}{Resulting Coating Structure} & 
\multicolumn{3}{c|}{$\sqrt{S^{\rm opt}_{\rm th}}$}
& 
$\sqrt{S^{\lambda/4}_{\rm th}}$ \\
 \cline{2-15}
$\phi_B/\phi_S$ & $N$ & 
\multicolumn{4}{c|}{First 4 layers} 
&
\multicolumn{2}{c|}{Repeated Pair}
&
\multicolumn{4}{c|}{Last 4 layers} 
&
$\frac{\phi_B}{\phi_S}=\frac{1}{5}$ & 
$\frac{\phi_B}{\phi_S}=1$ & 
$\frac{\phi_B}{\phi_S}=5$ & 
\\
\hline
1/5 & 42 & 
0.0479 & 0.1581 & 0.3430 & 0.1760 &  0.2919 & 0.1897 & 0.3164 & 
0.1738 & 0.3178 & 0.1627 & {\bf 5.01} & 6.64 & 8.81 & 5.35 \\
\hline
1 & 42 & 
0.1020 & 0.1250 & 0.3267 & 0.1917 & 0.2911 & 0.1914 & 0.3110 &  0.1752 & 0.3196 & 0.1609 & 5.02 & {\bf 6.64} & 8.81 &7.05 \\
\hline
5 & 42 & 
0.1118 & 0.0968 & 0.3353 & 0.1882 & 0.2893 & 0.1939 & 0.3135 & 
0.1673 & 0.3199 & 0.1662 
& 5.02 & 6.64 & {\bf 8.81} & 9.33  \\
\hline 
\end{tabular}
\caption{Results of coating-structure optimization. We list optimized coating structures for $T_{1064}=5\,$ppm and $T_{532}=5\%$, for three target values of $\phi_B/\phi_S$ while fixing the measured effective loss angle $\phi_D$ \R{[Eq.~\eqref{eqphiD}]} and other baseline material parameters [Table~\ref{tab2}]. Thickness of coating layers are given in units of wavelength (for 1064\,nm light). For each optimized coating structure,  thermal noise is calculated separately for all three values of $\phi_B/\phi_S$, and given in units of $10^{-21}\,\mathrm{m}/\sqrt{\rm Hz}$ (thermal noise for the target $\phi_B/\phi_S$ is given in boldface, and boldface numbers should be the minimum within its column); thermal noise spectra of the 38-layer $\lambda/4$ stack assuming the target $\phi_B/\phi_S$  are also listed for comparison.}
\label{tab:opt}
\end{table*}

 \begin{figure}[h]
  \centering
  \includegraphics[width=0.4\textwidth]{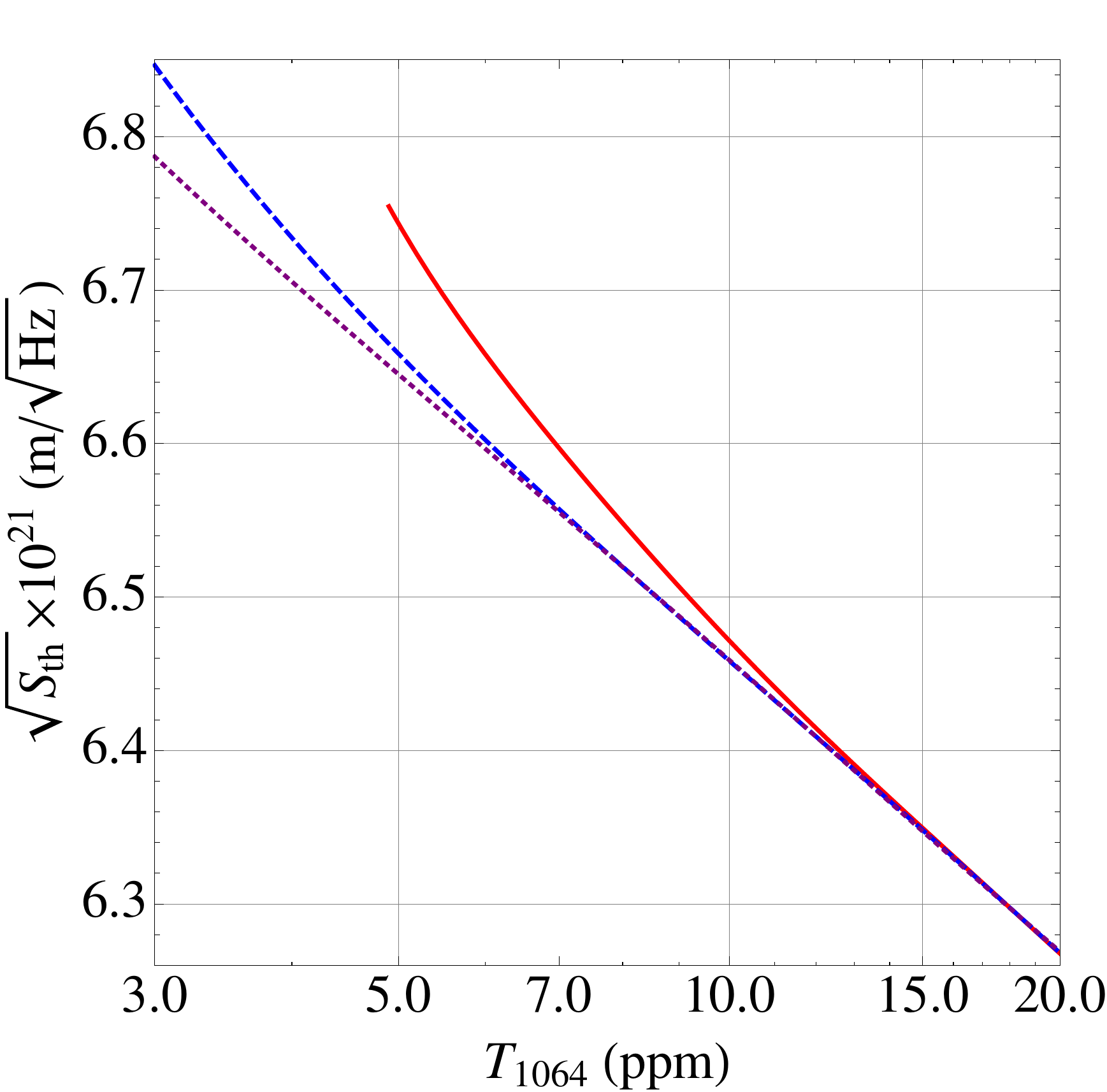}
  \caption{Optimized thermal noise versus transmissivity at 1064\,nm, for a coating of 
                 38 (red), 40 (blue), and 42 (purple) layers.}
\label{fig:TNvsT}
\end{figure}
 
Results for baseline material parameters (Table.~\ref{tab2}) and 
$N=$38, 40 and 42 have been shown in Figure~\ref{fig:TNvsT}. This figure indicates that different numbers of layers should be chosen for different target $T_{1064}$ -- more layers are required for lower transmissivity (higher reflectivity). Overall, the optimal thermal noise varies by around $\sim 10$\% as for $T_{1064}$ from 3 to 20\,ppm. In particular, for the standard Advanced LIGO requirement of 5\,ppm  (see first column of Table~\ref{tab:opt}),  42 layers are found to be optimal.  This is 2 more pairs or 4 more layers than the 38-layer $\lambda/4$ doublet, which has the minimum number of layers to reach 5\,ppm.  The larger number of layers here gets lower thermal noise (by 6\,\%) because the more lossy tantala layers are shortened, and the less lossy silica layers lengthened. 


We have further optimized the structure when the ratio $\phi_B/\phi_S$ is different from 1, while keeping fixed the effective loss angle measured so far --- as done in Sec.~\ref{subsec:lossangle}.  For $T_{1064}=5\,$ppm, we have listed results of optimized coating structure and thermal noise in the second and third columns of Table~\ref{tab:opt}. The extent of variation found here is comparable to those obtained in Sec.~\ref{subsec:lossangle} using  a standard coating structure without optimization: the optimal coating structures consistently lower thermal noise by about 6\%.  In addition, as shown  in  Table~\ref{tab:opt}, the optimal coating structure is robust against changes in $\phi_B/\phi_S$: structure obtained for any one of the values of the ratio is already almost optimal for all other ratios.


%
%

\section{Measurements of Loss Angles}
\label{sec:measurement}
In this section, we study possible mechanical ringdown experiments that can be used to measure 
independently the bulk and shear loss angles, $\phi_B$ and $\phi_S$  of a coating material. 

In a ringdown experiment, a sample with a high intrinsic Q is coated with a thin layer of the 
coating material in question.  Due to 
the mechanical losses in the coating, the quality factor of the mechanical eigenmodes of the sample will 
be reduced~\cite{glasgow, Penn}.  More specifically, 
for the $n^{th}$ eigenmode with resonant frequency $f_n$, if an 
e-folding decay time of $\tau_n$ is measured, then the quality factor is 
\begin{equation}
Q_n  = \pi f_n \tau_n\,,
\end{equation}
while correspondingly, the loss angle is given by
\begin{equation}
\phi(f_n) =1/Q_n\,,
\end{equation}
which is equal to the amount of energy dissipated $W_{\rm diss}$ \R{per radian. }

\begin{figure}[h]
     \centering
         \includegraphics[width=0.3\textwidth]{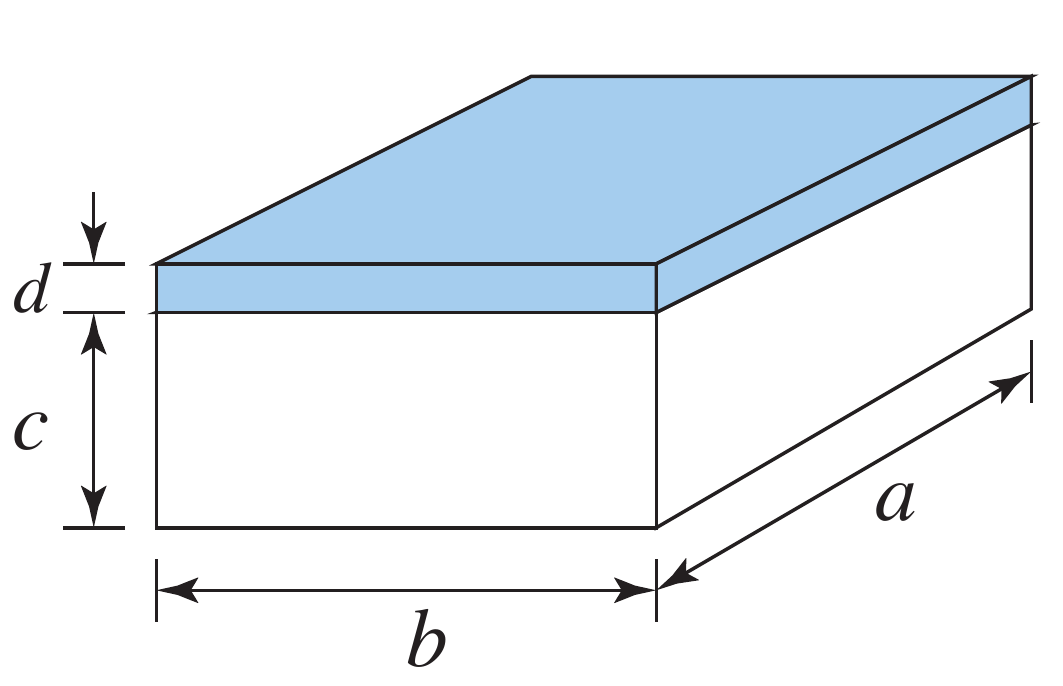}
                 \caption{Rectangular shaped thin plate ($a\times b\times c$) with thin coating (thickness d): $c\ll a,b; d\ll c$. 
                                The transverse vibration mode is considered in this case}
       \label{fig:rectangle}
\end{figure}

\subsection{Bending Modes of a Thin Rectangular Plate}
\label{subsec:bend}
Figure~\ref{fig:rectangle} shows the schematic geometry of a rectangularly shaped sample, 
in which a thin  coating layer with thickness $d$ is deposited on a rectangular plate with 
dimensions $a\times b\times c$  ($c\ll a, b$), and $d$ is much less than $c$. If we pay attention only 
to the transverse oscillations of the plate,  the amount of 
energy stored in the coating layer, in the form of bulk and shear energies $U_B$ 
and $U_S$, as a fraction of the entire energy $U$, can be calculated as \R{[Reference?]}
\begin{eqnarray}
\frac{ U_B }{U}&=&\frac{d}{c} \frac {Y_c}{Y_s} \frac{(1-\sigma_s^2)(1-2\sigma_c)}{(1-\sigma_c)^2} \\
\frac{ U_S}{U}&=&\frac{2d}{c} \frac {Y_c}{Y_s}\frac{(1-\sigma_s^2)(1-\sigma_c+\sigma_c^2)}{(1-\sigma_c)^2(1+\sigma_c)}
\end{eqnarray}
Using Eq.~\eqref{eq:totalangle}, the  the total loss angle of the sample is
\begin{align}
\phi &=\phi_{\rm sub}  \nonumber\\
&+ \frac{d}{c}\frac{Y_c}{Y_s}\frac{(1-\sigma_s^2)}{(1-\sigma_c)^2}\left[\phi_B(1-2\sigma_c)+2\phi_S\frac{1-\sigma_c+\sigma_c^2}{1+\sigma_c}\right].\quad
\label{eq:ringdownphi}
\end{align}

It is not surprising that this combination of $\phi_B$ and $\phi_S$ is proportional to $\phi_D$  [c.f.~Eq.~\eqref{eqphiD}], the loss angle of the 2-D flexural rigidity of the coating material, which we defined in Sec.~\ref{subsec:med}.  \R{This is because when the  drum mode of a thinly coated plate is excited, the stress $T_{zz}$ remains zero within the coating layer, and the layer's  elastic response  is governed by the flexural rigidity, as  defined  in Sec.~13 of Ref.~\cite{Landau}.} 

As it turns out, the part of coating thermal noise due to bending of the coating-substrate interface [$S_{z_s z_s}$ in Eq.~\eqref{sss}]  \R{also depends directly on $\phi_D$,  because the loss mechanism in this case is the same as during the oscillation of a drum mode --- one only applies a perpendicular force from below the coating layer, while keeping $T_{zz} =0$ within the layer.}

It proves less straightforward to connect the thickness fluctuation part of thermal noise  [$S_{u_z u_z}$ in Eq.~\eqref{sll}] to the effective loss angle of either $Y$ or $D$. Although the loss mechanism here is due to the compressing of a thin membrane from both sides --- this membrane is not characterized by vanishing $T_{xx}$ and $T_{yy}$, because the coating is attached to a substrate which provides restoring forces along the transverse ($x$ and $y$) directions.  However, in the case when the Poisson ratio $\sigma_c$ of the coating vanishes, the thickness fluctuation does depend on the loss angle of the Young's modulus. 

For our baseline parameters, \R{mechanical dissipation} is mostly contributed by the tantala layers, and because the Young's modulus of the tantala coating material is assumed to be much greater than that of the substrate, the largest contribution to the LIGO mirrors' Brownian noise is bending noise $S_{z_s z_s}$.  This explains why the noise only varies by 30\% (as noted in Sec.~\ref{subsec:lossangle}) even if no further measurements on the other loss angle is made.


\begin{figure}
    \centering
      \includegraphics[width=\columnwidth]{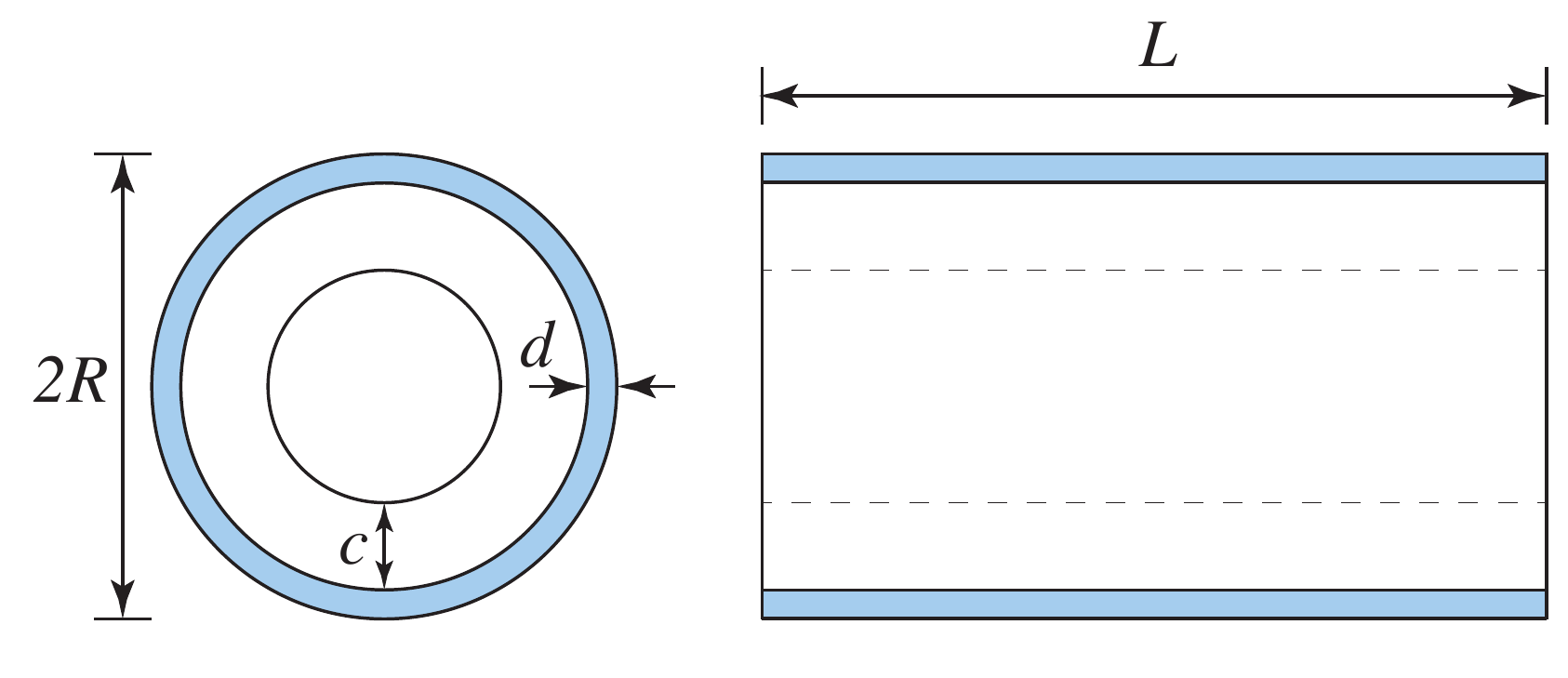}
        \caption{Thin cylindrical shell with thin coating outside. The first torsional eigenmodes of 
                      such a shell can be used to measure the shear loss angle of the coating.}
   \label{fig:cylinder}
\end{figure}

\subsection{Torsional Modes of a Coated Hollow Cylinder}
Here we propose an approach with which we can measure another 
combination of loss angles.  We consider a cylindrical shell with a thin, uniform coating layer outside, 
as shown in Fig.~\ref{fig:cylinder}($c\ll R$, $d\ll c$).   In this configuration, 
 the surface deformations produce strains in the plane of shell according to the Donnell shell theory ~\cite{Donnell}.
 Here we assumed that there is only  angular displacement in the shell, which means the longitudinal position of the cross section won't change. 
 For a torsion mode, we only have shear strain energy, the expressions are given by
\begin{eqnarray}
&&\frac{ U_B}{U}=0 \\
&&\frac{U_S}{U}=\frac{d}{c} \frac {Y_c}{Y_s} \frac{(1+\sigma_s)}{(1+\sigma_c)}\,.
\end{eqnarray}
As a consequence,  the total loss angle can be expressed as
\begin{eqnarray}
\phi=\phi_{\rm sub}+\frac{d}{c} \frac {Y_c}{Y_s} \frac{(1+\sigma_s)}{(1+\sigma_c)} \phi_S
\end{eqnarray}

For a cylinder shell, according to the Donnell shell theory, the natural frequency of the $n$-th torsional mode is given by~\cite{Blevins:Modes}
\begin{equation}
f_n=\frac{n}{2^{\frac{3}{2}}L}\left[\frac{Y}{\rho (1+\sigma)}\right]^{1/2}
\end{equation}

A more accurate calculation may be found by using the Fl\"{u}gge shell theory ~\cite{Flugge}.

Using the values from Table~\ref{tab:shell}, we can estimate the resonant frequency to be 
$9.2\,{\rm kHz}$.  The coating contribution to loss angle, assuming a $\phi_S$ of at least 
$10^{-5}$, would be at least the order of $10^{-6}$, which seems plausible to be extracted 
from ring-down measurements.

\begin{table}[h!]
\caption{Example parameters of a thin, uniformly coated cylindrical shell (${\rm SiO_2}$) \label{tab:shell}}
\begin{tabular}{l*{8}{c}r}
\toprule
&& $L$ && $R$ && $c$ && $d$\\
\hline
unit(mm)&& 200 && 50 && 1 && 0.04 \\
\hline
\end{tabular}
\end{table}

With the measurement of both the thin plate and cylinder shell, we can obtain $\phi_B$ and 
$\phi_S$ of the coating.

\section{Conclusions}

\label{sec:conclusions}

In this paper, by applying the Fluctuation-Dissipation Theorem, we obtained a 
{\it full set of  correlation functions} \eqref{sll}--\eqref{ssl} of Brownian thermal fluctuations 
of a multi-layer dielectric coating.  In particular, we have related  fluctuations of coating thickness 
and coating-substrate interface to {\it independent} bulk and shear thermal stresses  associated with  each coating layer.  These stresses not only \R{induce thickness} fluctuations of the layers themselves, but also bends the coating-substrate interface --- this bending noise had not been previously appreciated intuitively, although its effect has been incorporated into formulas, e.g., in Ref.~\cite{harry}.  As a result, we found that although 
thickness fluctuations of different coating layers are independent \R{of} each other, they each have partial correlations 
with the height fluctuations of the coating-substrate interface. 
Moreover, bulk loss creates a positive correlation between 
them, while shear loss creates a negative correlation.  The entire picture is succinctly written 
mathematically in Eqs.~\eqref{uzzeff} and \eqref{zeff}.  This coherence structure then gives coating 
Brownian noise in Eq.~\eqref{thermal_simple}. Apart from having provided a pedagogical and systematic derivation of these noise components,  the most important conceptual consequence of our work is to point out an uncertainty in coating loss angles, 
which has not been anticipated previously.  We have also incorporated the photo elastic effect, the  reflectivity fluctuations of the interfaces within the multilayer coating, and considered the effect of amplitude modulations caused by Brownian thermal noise. All of these turned out to be rather unimportant. 

\begin{table}
\begin{tabular}{c|c|c|c}
\hline\hline
\begin{tabular}{c}
material \\ parameter
\end{tabular} & range & \begin{tabular}{c} uncertainty \\ in $\sqrt{S_x}$ \end{tabular} & 
\begin{tabular}{c} 
for details,\\
see\end{tabular}   \\
\hline
$\phi_B/\phi_S$ & 0.2 -- 5\footnote{Fixing the combination $\phi_D$}  & $\pm 37$\% & Sec.~\ref{subsec:lossangle}, Figs.~\ref{fig:baseline_zeta}, \ref{fig:loss_total}.\\
\hline
$Y_{\rm Ta}$ &  factor of $\sim2$ & $\sim$60\% & Sec.~\ref{subsec:depyoung}, Fig.~\ref{fig:varyingYoungs}.\\
\hline
$\sigma_{\rm Ta}$ & $\pm$ 0.2  &  
\begin{tabular}{c} 
up to 10\% if  \\
$\phi_B/\phi_S \neq 1$
\end{tabular}
 & Sec.~\ref{subsec:depyoung}, Fig.~\ref{fig:varyingPoisson}. \\
 \hline
$\beta$ & $-1<\beta <+1$ & $\pm 1$\%~\footnote{Calculated from $\mathrm{Ta}_2\mathrm{O}_5$ layers} & Sec.~\ref{subsec:photoelastic}, Fig.~\ref{fig:photoelastic}. \\
\hline
\end{tabular}
\caption{Levels of thermal noise uncertainty caused by parameter uncertainties.\label{tab:uncertainty}}
\end{table}

We have applied our formalism to mirrors that are to be used in Advanced LIGO detectors.  As estimated in 
Sec.~\ref{sec:dependence} and summarized in Table~\ref{tab:uncertainty} (calculated for a typical candidate for the Advanced LIGO end test-mass mirror coating configuration),  parameter uncertainties could lead to non-negligible corrections to coating Brownian 
noise calculations.   The biggest uncertainties actually arise from the elastic moduli of coating materials --- for example, \B {current uncertainties in Young's modulus of the tantala coating material might lead up to $60\%$ increase in thermal noise}.  Although photo elastic coefficients for our coating materials are very uncertain, they do not significantly affect thermal noise since light does not penetrate through many layers. 

It is rather remarkable that our lack of experimental knowledge about the loss angles, beyond what we had already obtained from the ring down of drum modes, would not give rise to a higher uncertainty in thermal noise.  This is rather serendipitous, considering our path of understanding of the problem:  for the baseline parameters of Advanced LIGO, the highest contribution to coating Brownian noise arises from the coating-substrate bending noise caused by losses in tantala layers, because these layers are much more lossy than the silica layers, and have been assumed to have a much higher Young's modulus than the substrate material.  This bending noise, first elaborated by this work, turns out to be associated with the loss angle of the 2-D flexural rigidity, which in turn is directly connected to the decay of the drum modes of a thinly coated sample.  This means the currently existing program has been measuring the predominant loss angle all along.    Nevertheless, the level of uncertainty noted in our study still warrants further experiments seeking the other loss angle, e.g., as outlined in  Sec.~\ref{sec:measurement}.  In addition, since future gravitational-wave detectors may use different substrate and coating materials, situations may arise when the loss angle measured  now does not correlate with the total coating brownian noise.

\R{At this moment, it is worth looking once more at the previously used loss angles, $\phi_\parallel$ and $\phi_\perp$ --- although they are mathematically ill defined, they do correctly reflect the existence of two channels of loss.   The $\phi_\parallel$ was meant to characterize losses incurred by the $x$-$y$ deformations of the coating measurable when we do not compress the coating but instead drive its deformations using drum modes of the substrate.  This loss angle is now replaced by the (mathematically well-defined) imaginary part of the flexural rigidity, for which extensive measurements have already been carried out.  The $\phi_\perp$ was meant to characterize the losses incurred by compressing the coating layers.  This has not been measured because it had not been obvious how to easily excite this mode of coating deformation (the most obvious way would be to compress the coating layer, but that is difficult); however, because the Young's modulus of the coating is much larger than that of the substrate, this difficult-to-measure loss angle should not contribute as much to the total coating noise.  This said, in this work, we do come up with ways to measure both loss angles, $\phi_S$ and $\phi_B$, without having to compress the coating layers --- but instead by exciting different modes of substrate deformation.  Of course, this is only possible because we have assumed that the material is isotropic --- otherwise we may have to compress the coating to directly access the loss induced by such a deformation. }

On the other hand, one may think of the possibilities of using substrate materials with higher Young's modulus to reduce the bending noise. Sapphire and Silicon are two viable choices because they both have higher Young's modulus than tantala. Using Eq. \eqref{sll}--\eqref{ssl}, it is straight forward to estimate the new coating brownian noise while replacing the substrate material by sapphire or silicon but keeping the same aLIGO coating design. It turns out that the coating brownian noise will be reduced to $35\%$ of its original power spectra value if we use silicon substrate or $32\%$ if we use sapphire. However, there are other disadvantages for sapphire or silicon substrate that prevents us from using them for aLIGO mirrors. The main problem is that they both have very high thermal conductivity - much higher than fused silica. As a result, the substrate thermoelastic noise is one of the important noise source for both materials. For instance, if the aLIGO mirror was made of sapphire, the bulk thermal elastic noise would have about the same magnitude as the coating brownian noise at 100 Hz. As for silicon substrate, the bulk thermal elastic noise is more than 4 times larger than its corresponding coating brownian noise in power because silicon has even higher thermal conductivity than sapphire. One may refer to \cite{braginsky} for detailed methods to calculate bulk thermal elastic noise. Setting up the experiment in a cryogenic enviroment is a possible way to reduce the thermooptic noise.

Furthermore, our formula Eq.~\eqref{thermal_simple} can serve as a starting point for optimizing the material choice and structure design of the multi-layer coating taking light penetration effects into account.  Our numerical results in Sec.~\ref{sec:optimization} (see Table~\ref{tab:opt}) have shown that optimization of the coating structure consistently offers a $\sim 6$\% decrease in thermal noise, regardless of $\phi_B/\phi_S$.  In fact, the optimal structure for these ratios are quite similar, and configurations obtained for each presumed ratio of $\phi_B/\phi_S$ are shown to work for other ratios interchangeably.  

Upon completion of this 
manuscript, we noted that the optimization of coating structure for the case assuming $\phi_B = \phi_S$ (and $\beta =0$) has been carried out by Kondratiev, Gurkovsky and Gorodetsky~\cite{Gorodetsky}. [We note that their formalism is capable to treating $\beta\neq 0$ and $\phi_B\neq \phi_S$, as well as back-scattering induced by photo elasticity, but they did not explore the impact of these effects in their optimization.]  Our results are compatible with theirs, if we also use these restrictions in parameter space and ignore back-scattering. 

A comparison between our result, Kondratiev et al., and Harry et al.~\cite{harry} (which ignores light penetration into the layers, and also effectively assumes $\phi_S =\phi_B$) would therefore illustrate the effects caused by ignoring photoelasticity and further ignoring light penetration into the coating. This is shown in Table~\ref{tab:tncompare}.  This again confirms that for total coating thermal noise, light penetration causes noticeable difference in coating thermal noise, while photoelasticity causes a negligible difference.

\begin{table*}
\begin{tabular}{c|c|c|c}
\toprule
Coating & \begin{tabular}{c} 
Ref.~\cite{harry} \\
(no light penetration)
\end{tabular}
& \begin{tabular}{c} 
Ref.~\cite{Gorodetsky}\\
($\beta=0$ and no back scattering) 
\end{tabular}
& This Work \\
\hline
$\lambda/4$ & 7.18 & 7.08 & 7.08 \\
\hline
Advanced LIGO & 6.93 & 6.82 & 6.83 \\
\hline
optimal & 6.73 & 6.62 & 6.64 \\
\hline
\end{tabular} 
   \caption{Comparison of thermal noise spectral density (assuming $\phi_B=\phi_S$ and evaluated 
                  at 200\,Hz, in units of $10^{-21}{\rm m}/\sqrt{{\rm Hz}}$)  between different works.
\label{tab:tncompare}}
\end{table*}


\begin{acknowledgments}
We would like to thank Stan Whitcomb, Raffaele Flaminio, Jan Harms, 
Gregg Harry, Yasushi Mino, Valery Mitrofanov, Kentaro Somiya, Sergey Vyatchanin, and 
other members of the LSC Optics Working Group for very useful discussions. 
We thank Iain Martin and Andri Gretarsson  for many useful suggestions to the 
manuscript. This work was 
supported by NSF Grant PHY-0555406, PHY-1068881 and CAREER Grant PHY-0956189, 
the David and Barbara Groce Startup Fund, and the David and 
Barbara Research Assistantship at the California Institute of Technology.  Funding has also been 
provided by the Institute for Quantum
Information and Matter, an NSF Physics Frontiers Center with support of
the Gordon and Betty Moore Foundation.
\end{acknowledgments}

\appendix
\section{Fluctuations of the Complex Reflectivity due to Refractive index fluctuations}
\label{app:n_fluc}


\R{Brownian noise is not only caused by the random strains, but also by the refractive-index fluctuations caused by such strains, through the photo elastic effect [Cf.~Eqs.~\eqref{coth} and \eqref{eq:dn}].  We shall quantify this  contribution in this section.}

\subsection{The photoelastic effect}
\label{photoelastic}
 If we denote the displacement of coating mass elements as $(u_x,u_y,u_z)$, then the relative 
coating-thickness change from its equilibrium value  can be written as
\begin{equation}
\delta l/l = u_{z,z}
\end{equation}
and the relative transverse area expansion can be written as
\begin{equation}
\delta A/A = u_{x,x} + u_{y,y}
\end{equation}
If we denote 2-dimensional displacement vectors along the $x$-$y$ plane as $\vec u=(u_x,u_y)$, 
and two-dimensional gradient as $\vec\nabla$, then we have
\begin{equation}
\delta A/A = u_{x,x} + u_{y,y} =\vec\nabla\cdot\vec u
\end{equation}
We can then write the change in refractive index as
\be
\label{dn_terms}
\delta n=\left[\frac{\partial n}{\partial \log l}\right]_{A_j} \frac{\delta l}{l}+\left[\frac{\partial n}{\partial \log A}\right]_{l_j}\vec\nabla\cdot\vec u
\ee
where $\partial n/\partial \log l$ and $\partial n/\partial \log A$ only \R{depends on material properties.}
The two terms on the right-hand side of Eq.~\eqref{dn_terms} represent refractive index change driven 
by relative length and area changes, respectively. The {\it first term} is given by~\cite{stone} 
\be
\label{betaL}
\beta^L=\left[\frac{\partial n}{\partial \log l}\right]_{A}=-\frac{1}{2}n^3CY
\ee
where $C$ is the photoelastic stress constant, $Y$ is the Young's modulus. For silica, $CY \approx 0.27$, therefore $\rm\beta_{Si}^L=-0.41$. The photoelastic coefficient can also be written as
\be
\beta =-\frac{1}{2}n^3p_{ij}
\ee
where $p_{ij}$ is the photo elastic tensor~\cite{lax}. Some experiments have been done to measure this 
coefficient for tantala~\cite{Nakagawa}. Empirically, the value of $p_{ij}$
varies  from $-0.15$ to 0.45 for ${\rm Ta_2O_5}$ thin film fabricated in different ways. Here for the longitudinal 
photoelasticity, $\rm\beta_{Tan}^L$ , we use $-0.5$ in our numerical calculation.


  We shall next obtain formulas that will allow us to convert 
fluctuations in $n$ into fluctuations in the complex reflectivity of the multi-layer coating.

\subsection{Fluctuations in an Infinitesimally thin layer}
\label{app:inf:layer}

\R{Because the coating is much thinner than the beam spot size, we only consider phase shifts along the $z$ direction --- for 
each value of $\vec x$.  }
If the refractive index $\delta n$ at a particular location $\delta n(z)$ is 
driven by longitudinal strain $u_{zz}$ at that location, the fact that 
$\langle u_{zz} (z') u_{zz}(z'')\rangle \propto \delta(z'-z'')$ causes concern, because this indicates 
a high {\it variance} of $\delta n$ at any given single point $z$, with a magnitude which is formally 
infinity, and in reality must be described by additional physics (for example, there would be a scale 
at which the above-mentioned delta function starts to  become resolved).  Therefore, if we naively 
considers the reflection of light across an interface, at $z=z_0$, then the independent and 
high-magnitude fluctuations of $n(z_0-)$ and $n(z_0+)$ would lead to a dramatic fluctuation in 
the reflectivity 
\begin{equation}
r =\frac{n(z_0-)-n(z_0+)}{n(z_0+)+n(z_0-)}
\end{equation}
\R{of the interface, whose magnitude of fluctuation seems to be indefinitely large.  Fortunately, for any thin layer, if we simultaneously consider  
propagation through this layer and the reflection and transmission across {\it both} of its boundaries, 
then the effect caused by the refractive index fluctuation of this particular layer can be dramatically 
suppressed.   Nevertheless, we do find an additional fluctuating contribution to the total complex 
reflectivity of the multi-layer coating.  }

\begin{figure}
\includegraphics[width=0.2\textwidth]{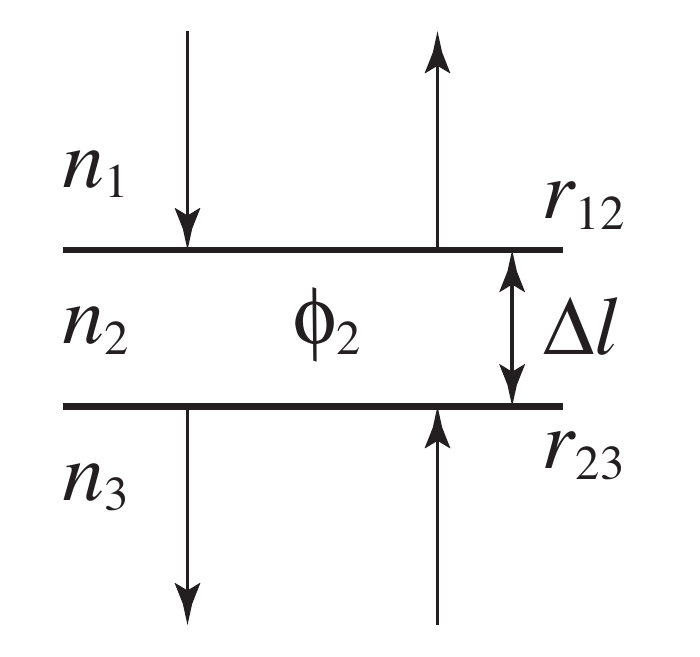}
\caption{Light propagation across a thin layer (thickness of $\Delta l$) with fluctuating refractive index (from a uniform $n_2$ to an average of $n_2+\delta n_2$ within this thin layer).  The propagation matrix corresponding to this structure is given by Eq.~\eqref{eqMlayer}. \label{fig_thin} }
\end{figure}

In order to carry out \R{a correct calculation that does not diverge, we first consider a three-layer and two-interface 
situation, as shown in Fig.~\ref{fig_thin}, with $n_1$, $n_2$ and $n_3$ separated by two interfaces, with the length of the $n_2$ 
layer given by $\Delta l$ --- and here we only consider fluctuations in $n_2$.  The entire transfer 
matrix (from below to above, in Fig.~\ref{fig_thin}) is given by }
\begin{equation}
\label{eqMlayer}
\mathbf{M} = \mathbf{R}_{r_{12}} \mathbf{T}_{\phi_2} \mathbf{R}_{r_{23}}
\end{equation}
following the same convention as in Sec.~\ref{sec:fluctuation}. Suppose the originally uniform $n_2 $ now fluctuates, and after averaging over this think layer, gives a mean refractive index of $n_2 + \delta n_2$, we use this as the refractive index of the entire layer, and then have
\begin{equation}
\delta{\mathbf{M}} = \frac{   n_2}{\sqrt{n_1 n_3}}\left(
\begin{array}{cc}
i & -i \\
i  & -i
\end{array} 
\right)  \delta n_2 \cdot k_0 \Delta l
\label{dM}
\end{equation}

This \R{can be considered as a regularization, because each individual} $\mathbf{R}_{r_{12}}$ or 
$\mathbf{R}_{r_{12}}$ (since their expressions only contain $n_1$, $n_2$ and $n_3$ but no $l$) has a 
standard deviation proportional to $O(1/\sqrt{\Delta l})$ (when $\Delta l$ is greater than the coherence length of refractive-index fluctuation) or $O(1/\Delta l)$ (when $\Delta l$ is less than the coherence length of refractive-index fluctuation) --- both diverge as $\Delta l \rightarrow 0$ --- \R{which means the reflectivity fluctuation of each of these layers diverge.}  However, in order for our use of average refractive index to make sense in calculating the reflectivities $r_{12}$ and $r_{23}$, $\Delta l$ should be less than the coherence length of refractive index fluctuations.  In any case, the total transfer matrix $\delta M$ does not diverge; it instead has an infinitesimal fluctuation.  Moreover, since $\delta M$ only depends on $\delta n_2 \cdot \Delta l$, we shall see that the particular choice of $\Delta l$ will not affect the final results when layers like these are stacked together.

The physical meaning of Eq.~\eqref{dM}  is clear: a random field of refractive index not only gives a 
random phase shift (diagonal term), but also gives rise to a random reflectivity (non-diagonal term).  
The latter term is an additional contribution that has been ignored by previous calculations.

\subsection{The entire coating stack}
\label{app:stack}
Now we are ready to consider the entire multi-layer coating.  Here we bear in mind that eventually, the fluctuation in $n$ has a non-zero coherence length --- and we can then divide our existing layers further into sub layers with length $\delta l$ much less than the physical coherence length.  Since each of these sub layers only makes a negligible contribution to the entire complex reflectivity, we only need to consider \R{layers that contain only one coating material}. Let us first focus on Layer $j$, bounded by two interfaces with reflectivities $r_{j-1}$ and $r_j$, respectively.  The total transfer matrix of the entire stack is written as
\begin{equation}
\mathbf{M} = \cdots  \mathbf{T}_{\phi_{j+1}} \mathbf{R}_{r_j} \mathbf{T}_{\phi_j} \mathbf{R}_{r_{j-1}} \cdots
\end{equation}
Here reflectivity fluctuations within Layer $j$ is going to add to the matrix $\mathbf{T}_{\phi_j}$ above. Consider $dz$-thick sub-layer located at distance $z'$ from the $r_{j}$ boundary (lower boundary in Fig.~\ref{fig:coating_config}), therefore at coordinate location $z=z_{j+1}+z'$ and integrate, we have
\begin{align}
\mathbf{T}_{\phi_j} &\rightarrow \mathbf{T}_{\phi_j}+ k_0 
\int_0^{l_j}  
\delta n (z_{j+1}+z)
\mathbf{T}_{k_0 n_j z} 
\left[
\begin{array}{cc}
i & -i \\
i  & -i
\end{array} 
\right]      
\mathbf{T}_{k_0 n_j (l_j- z)} \, dz' 
\nonumber\\
&= 
\left[
\begin{array}{cc}
1   &  \delta\eta_j \\
 \delta\eta_j^* & 1 
\end{array}
\right] \mathbf{T}_{\phi_j+k_0 \delta \bar n_j l_j}  
\label{newT}
\end{align}
where
\begin{eqnarray}
\delta\bar n_j  &=&\frac{1}{l_j}\int_0^{l_j}\delta n_j(z_{j+1}+z) dz
\end{eqnarray}
and 
\begin{eqnarray}
\delta \eta_j &=& -i k_0 \int\delta n_j (z_{j+1}+z)e^{2 i k_0 n_j z} dz
\end{eqnarray}
Here we have defined 
\begin{equation}
z_{j} \equiv \sum_{n=j}^{N} l_n\,.
\end{equation}
to be the $z$ coordinate of the top surface of Layer $j$.

We need to adapt the new transfer matrix into the old form, but with modified $\{r_j\}$ and $\{\phi_j\}$. From Eq.~\eqref{newT}, since $\delta\eta_j$ is complex, we need to adjust $\phi_j$, $r_j$, as well as $\phi_{j+1}$:
\begin{eqnarray}
&&\mathbf{T}_{\phi_{j+1}} \mathbf{R}_{r_j}  \mathbf{T}_{\phi_j}\nonumber\\
& \rightarrow &
\mathbf{T}_{\phi_{j+1}+\delta\psi_{j}^+} \mathbf{R}_{r_j +\delta r_j}  \mathbf{T}_{\phi_j +k_0 l_j \delta \bar n_j +\delta\psi_j^-}
\end{eqnarray}
Here we have defined in addition
\begin{equation}
\label{dr}
\delta r_j  = -t_j^2 k_0\int_0^{l_j} \delta n_j (z_{j+1}+z) \sin( 2 k_0 n_j z)dz
\end{equation}
and
\begin{equation}
\delta\psi_j^{\pm} =\frac{r_j^2 \pm 1}{2r_j} k_0 \int_0^{l_j} \delta n_j (z_{j+1} +z) \cos(2 k_0 n_j z) dz
\end{equation}
As we consider photoelastic noise of all the layers together, $\delta r_j$ in Eq.~\eqref{dr} needs to be used for the effective fluctuation in reflectivity of each layer, while 
\begin{equation}
\label{dphijapp}
\delta \phi_j = k_0 l_j \delta\bar n_j + \delta\psi_j^- + \delta \psi_{j-1}^+
\end{equation}
should be used as the total fluctuation in the phase shift of each layer.

\subsection{Unimportance of transverse fluctuations}
\label{app:transverse}

Connecting with photoelastic effect, we have explicitly
\begin{equation}
\delta n_j (z,\vec x) =\beta_j^L u_{zz}(z,\vec x) + \beta_j^T \vec\nabla \cdot \vec u
\end{equation}
Here the vector $\vec u$ is the two-dimensional displacement vector $(u_x,u_y)$ and $\vec \nabla \cdot $ is the 2-D divergence along the $x$-$y$ plane.  For terms  that contain $\vec u$, we note that when taking the optical mode into account [see Sec.~\ref{phasethermalnoise}], i.e., when a weighted average of $\xi$ is taken, they yield the following type of contribution
\begin{eqnarray}
&&\int_M I(\vec x) \left(\vec \nabla\cdot \vec u\right) d^2\vec x   \nonumber\\ 
&=&\int_{\partial M} dl (\vec n \cdot \vec u I) + \int_M \vec u\cdot\vec \nabla I \,d^2 \vec x \nonumber\\
&=&\int_M \vec u\cdot\vec \nabla I \,d^2 \vec x
\end{eqnarray} 
Here $M$ stands for the 2-d region occupied by the beam, and $\partial M$ is the boundary on which power already vanishes. As a consequence,
the first term is zero according to the boundary condition, while the second term gains  a factor of $(l_i/r_{\rm beam})$ with respect to other types of coating Brownian noise, here $l_j$ is the thickness of the $j$-th layer, and $r_{\rm beam}$ is an effective beam radius.   Since we always assume coating thickness $l_i$ to be much smaller than the beam radius $r_{\rm beam}$, we can neglect refractive index fluctuation due to area fluctuation.

\section{Elastic Deformations In The Coating}
\label{app:elasticity}

Throughout this paper, we assume the mirror substrate to be a half infinite space.  We establish a Cartesian coordinate system, with $(x,y)$ directions along the coating-substrate interface, and $z$ direction orthogonal to the mirror surface (in the elasticity problem, we also ignore mirror curvature).   This allows us to calculate elastic deformations in the spatial frequency domain.  We will also assume the coating thickness to be much less than the beam spot size. 

We denote the displacement along $x$, $y$ and $z$ directions as $u_x$, $u_y$ and $u_z$. It is then straightforward to express the $3\times 3$ strain tensor $\mathbf{S}$ in terms of their derivatives, and stress tensor $\mathbf{T}$ in terms of Hooke's Law:
\ba
S_{ij}&=&\frac{1}{2}  \left(\frac{\partial u_i}{\partial x_j}+\frac{\partial u_j}{\partial x_i} \right) \\
\Theta &=&S_{ii} \\
\Sigma_{ij}&=&\frac{1}{2}\left[S_{ij}+S_{ji}\right]-\frac{1}{3}\delta_{ij} \Theta \\ 
T_{ij}&=& -K \Theta I_{ij}-2\mu \Sigma_{ij}\,.
\ea
Here we have $x^j=(x,y,z)$, with Latin indices (like $i$ and $j$) running from 1 to 3. Within any layer, it is straightforward to write down the most general  solution of the elasticity equilibrium equation
\begin{equation}
T_{ij,j}=0
\end{equation}
as
\begin{eqnarray}
\label{gen1}
\tilde u_x &=& i k_x[ (\tilde \alpha_+  + \kappa z \tilde \beta_+) e^{\kappa z}
+
 (\tilde \alpha_- - \kappa z \tilde \beta_-) e^{-\kappa z}
] \nonumber \\
&-& i k_y [\tilde \gamma_+ e^{\kappa z} +\tilde \gamma_- e^{-\kappa z}]
\\
\tilde u_y &=& i k_y[ (\tilde \alpha_+  + \kappa z \tilde \beta_+) e^{\kappa z}
+
 (\tilde \alpha_- - \kappa z \tilde \beta_-) e^{-\kappa z}
] \nonumber \\
&+ & i k_x [\tilde \gamma_+ e^{\kappa z} +\tilde \gamma_- e^{-\kappa z}]\\
\label{gen2}
\tilde u_z &=& -\kappa [\tilde\alpha_+ +\tilde\beta_+ (-3 +4\sigma+\kappa z)]e^{\kappa z} \nonumber \\
&+&\kappa [\tilde\alpha_- +\tilde\beta_- (-3+4\sigma-\kappa z)]e^{-\kappa z}
\end{eqnarray}
where tilde denotes quantities in the $x$-$y$ spatial-frequency domain, and $\kappa=\sqrt{k_x^2+k_y^2}$. Namely
\begin{equation}
u_x(x,y,z)=\int\frac{dk_x dk_y}{(2\pi)^2} \tilde u(k_x,k_y,z) e^{-i (k_x x+k_y y)}
\end{equation}

\begin{figure}[t]
\centering
\includegraphics[width=0.45\textwidth]{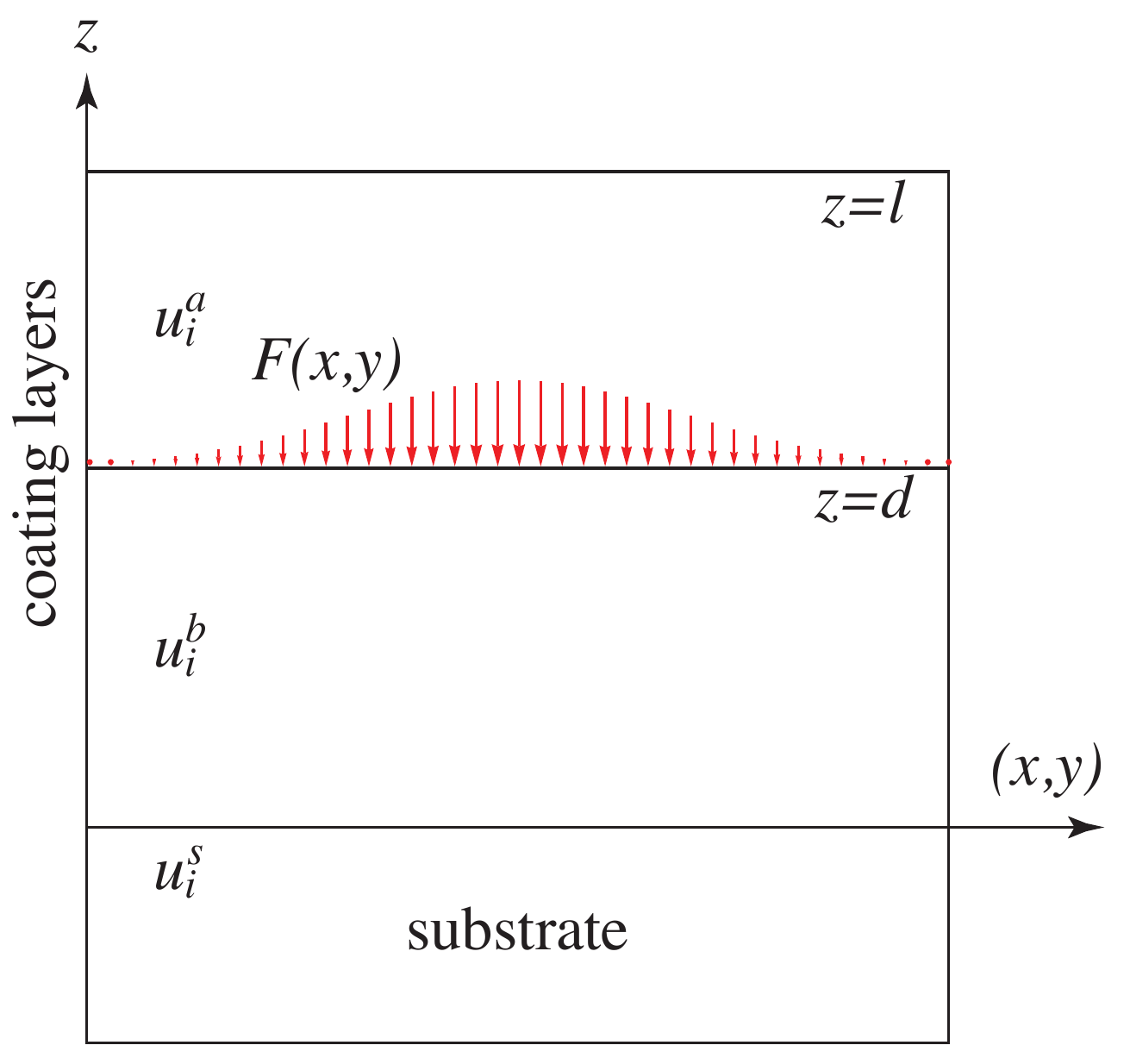}
\caption{Sample with single layer coating, force is applied perpendicular to the air/coating interface. \label{single}}
\end{figure}

We now consider a single-layer coating  on a substrate, with the coating-substrate interface located at $z=0$, and the coating-air interface at $z=l$.  Suppose there is a force profile  $F(x,y)$ exerted perpendicular to the coating surface, at $z=d$, $0<d<l$, and let us calculate the elastic deformation field caused by $F$.  The entire system is now divided into three regions, (a): $d<z<l$, (b): $0<z<d$, and (s): $z<0$. At the interfaces, we obtain the following 15 boundary conditions,
\ba
T^a_{iz}=0\,,&& \!\!\!\!\!z=l \;\\
T^a_{xz}=T^b_{xz}\,,\; T^a_{yz}=T^b_{yz}\,,\;
T^b_{zz}-T^a_{zz}=F\,, && \!\!\!\!\!z=d \;\\
u^a_j =u^b_j \,,&&\!\!\!\!\! z=d\quad \\
T^b_{iz}=0\,,\; u^b_j =u^s_j\,, && \!\!\!\!\!z=0 \;
\ea
as well as the condition that when $z\rightarrow -\infty$, $u_j^s\rightarrow 0$ (which leads to $\tilde\alpha_-^s  =\tilde\beta_-^s=\tilde\gamma_-^s=0$).   We are left with 15 fields of 
\begin{equation}
(\tilde\alpha^a_\pm,\tilde\beta^a_\pm,\tilde\gamma^a_\pm,\tilde\alpha^b_\pm,\tilde\beta^b_\pm,\tilde\gamma^b_\pm,\tilde\alpha^s_+,\tilde\beta^s_+,\tilde\gamma^s_+)
\end{equation}
which can be solved from the 15 boundary conditions. Assuming $\kappa d\ll1$ and $\kappa l\ll 1$, we obtain that all $\tilde \gamma$ vanish, and 
\ba
&&\tilde\alpha_+^a=\frac{F(1+\sigma_s)[2-3\sigma_s+\sigma_c(-3+4\sigma_s)]}{2Y_s\kappa^2(-1+\sigma_c)} \\
&&\tilde\alpha_-^a =\frac{F(\sigma_c-\sigma_s)(1+\sigma_s)}{2Y_s\kappa^2(-1+\sigma_c)}\\
&&\tilde\beta_+^a=-\frac{F(1+\sigma_s)(-3+4\sigma_s)}{4Y_s\kappa^2(-1+\sigma_c)} \\
&&\tilde\beta_-^a =\frac{F(1+\sigma_s)}{4Y_s\kappa^2(1-\sigma_c)}\\
&&\tilde\alpha_+^b=\frac{F(1+\sigma_s)[2-3\sigma_s+\sigma_c(-3+4\sigma_s)]}{2Y_s\kappa^2(-1+\sigma_c)}\\
&&\tilde\alpha_-^b=\frac{F(\sigma_c-\sigma_s)(1+\sigma_s)}{2Y_s\kappa^2(-1+\sigma_c)}\\
&&\tilde\beta_+^b=\frac{F[Y_s(1+\sigma)-Y_c(-3+\sigma_s+4\sigma_s^2)]}{4Y Y_s\kappa^2(-1+\sigma_c)}\\
&&\tilde\beta_-^b=\frac{F[Y_s(1+\sigma_c)-Y_c(1+\sigma_s)]}{4Y Y_s \kappa^2(-1+\sigma_c)}\\
&&\tilde\alpha_+^s=\frac{F(1+\sigma_s)(-1+2\sigma_s)}{Y_s \kappa^2}\\
&&\tilde\beta_+^s=-\frac{F(1+\sigma_s)}{Y_s\kappa^2}
\ea
We can therefore obtain the stain tensor in the frequency domain for the coating, the non-zero elements for region (a) are given by
\ba
\label{sxxa}
&&S_{xx}^a= \frac{Fk_x^2(-1+2\sigma_s)(1+\sigma_s^2)}{Y_s\kappa^2}\\
&&S_{yy}^a=\frac{Fk_y^2(-1+2\sigma_s)(1+\sigma_s^2)}{Y_s\kappa^2}\\
&&S_{xy}^a=S_{yx}=\frac{F k_x k_y(-1+2\sigma_s)(1+\sigma_s^2)}{Y_s\kappa^2}\\
&&S_{zz}^a=F\frac{\sigma_c(-1+\sigma_s+2\sigma_s^2)}{ Y_s(-1+\sigma_c)}
\ea
while those in  region (b) are given by
\ba
&&S_{xx}^b= \frac{Fk_x^2(-1+2\sigma_s)(1+\sigma_s^2)}{Y_s\kappa^2}\\
&&S_{yy}^b=\frac{Fk_y^2(-1+2\sigma_s)(1+\sigma_s^2)}{Y_s\kappa^2}\\
&&S_{xy}^b=S_{yx}=\frac{F k_x k_y(-1+2\sigma_s)(1+\sigma_s^2)}{Y_s\kappa^2}\\
&&S_{zz}^b=F\left[\frac{-(1+2\sigma_c)}{Y_c}-\frac{\sigma_c(-1+\sigma_s+2\sigma_s^2)}{ Y_s(1-\sigma_c)}\right]\quad
\label{szzb}
\ea
Using linear superposition, as well as taking the appropriate limits of the above solution, 
\R{it is straightforward to obtain elastic deformations in all the setups required in 
order to obtain cross spectra between different noises.}

\section{Definition of loss angle}
\label{App:lossangle}
In the past~\cite{harry},  the coating loss angle was defined in association with the 
parallel and perpendicular coating strains. The equation is written as
\begin{eqnarray}
\phi_{\rm coated}=\phi_{\rm sub}+\frac{\delta U_\parallel d}{U}\phi_\parallel+\frac{\delta U_\perp d}{U} \phi_\perp
\end{eqnarray}
where $\delta U_\parallel$ and $\delta U_\perp$ are the dissipation energy density in parallel and 
perpendicular coating strains
\begin{eqnarray}
&&\delta U_\parallel=\int_s \frac{1}{2}(S_{xx}T_{xx}+S_{yy}T_{yy})\,dx dy\\
&&\delta U_\perp=\int_s \frac{1}{2}S_{zz}T_{zz}\,dx dy
\end{eqnarray}
and where $S_{ij}$ are the strains and $T_{ij}$ are the stresses. While such a definition seems to be 
compatible with the symmetry of the system, the quantities $\delta U_\parallel$ and $\delta U_\perp$ 
cannot be used as energy, since in certain scenarios they each can become negative.

\begin{figure}[h]
\centering
\includegraphics[width=0.25\textwidth]{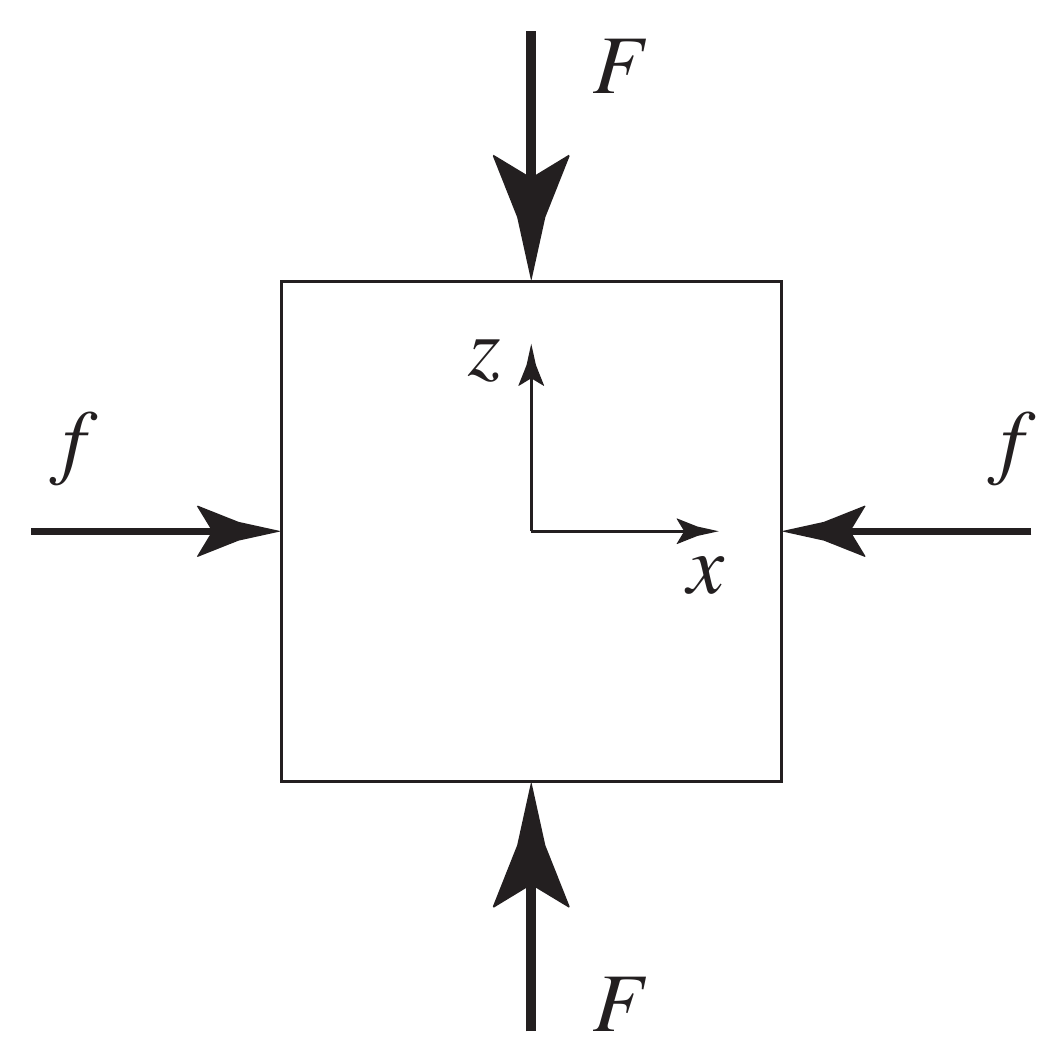}
\caption{Solid cube with two pairs of forces applied on the side: $f\ll F$ \label{cube}.}
\end{figure}

For example, if we have a cube with surface area of each side A (poisson ratio $\sigma$, Young's 
modulus Y), and we uniformly apply two pairs of forces, one pair with magnitude $f$ on opposite 
$yz$ planes, the other with magnitude $F$ on opposite $xy$ planes, with $f \ll F$, as shown in 
Figure~\ref{cube}. According to definition of Young's modulus and Poisson's ratio, up to leading order 
in $f/F$ the non-vanishing strains are, 
\begin{equation}
S_{zz}=-\frac{F/A}{Y}, \quad  S_{xx}=S_{yy}=\sigma \frac{F/A}{Y}
\end{equation}
On the other hand, for stress, we have, up to leading order in $f/F$, 
\begin{equation}
T_{xx} = -f/A\,,\quad T_{yy}=0\,,\quad T_{zz} =- F/A\,.
\end{equation}
As a consequence, we have
\begin{equation}
\delta U_\parallel =  S_{xx} T_{xx}+ S_{yy} T_{yy} =-\sigma f F/(A^2 Y) <0
\end{equation}
which means $\delta U_\parallel$ is not a reasonable candidate for energy, at least with $\sigma \neq 0$.  
Since it is also true that $S_{xx} T_{xx}<0$ we will arrive at 
\begin{equation}
\delta U_\perp = S_{zz}T_{zz}  < 0 
\end{equation}
if we take this configuration and rotate for 90 degrees around the $y$ axis, such that $x$ rotates into $z$.

One reasonable way of defining the loss angle is to derive from the fundamental elastic energy equation. 
The general form of the stored elastic energy density $U$ can be written as
\begin{eqnarray}
&&U=\frac{1}{2}K\Theta^2+\mu\Sigma_{ij}\Sigma_{ij}\\
&&U_B=\frac{1}{2}K\Theta^2\\
&&U_S=\mu\Sigma_{ij}\Sigma_{ij}
\end{eqnarray}
Where $K$ is called the {\it bulk modulus} and $\mu$ is the {\it shear modulus}. In the calculation, 
we use Young's modulus $Y$ and Poisson's ratio $\sigma$ instead of $K$ and $\mu$. Their relation 
are given in Eq.(15). The expansion $\Theta$ and shear $\Sigma$ are both irreducible tensorial parts of 
the strain tensor {\bf $S$}.
\begin{eqnarray}
&&\Theta=S_{ii}\\
&&\Sigma=\frac{1}{2}(S_{ij}+S_{ji})-\frac{1}{3}g_{ij}S_{kk}
\end{eqnarray}
Note that the expansion and shear energy $U_B$ and $U_S$ is always positive, so it is consistent to 
define loss angle by $\phi_B$ and $\phi_S$.

\section{Advanced LIGO style coating  }
\label{app:adv}

\begin{table}[h]
\begin{tabular}{c |ccccc}
\toprule
$j$ & \multicolumn{5}{c}{$l_j$} \\
\hline
1--5 & $0.497325$ & $0.208175$ & $ 0.289623$ & $0.237274$& 0.250176\\
\hline
6--10& 0.245330 & 0.249806 & 0.240129 & 0.270968 & 0.224129\\
\hline
11--15 & 0.251081 & 0.259888 &  0.260826 & 0.213460 & 0.290468\\
\hline
16--20 &  0.214524 & 0.273240 & 0.230905 & 0.259924 & 0.230020\\
 \hline
21--25 &  0.275429 &  0.233086 & 0.270385 & 0.208581 & 0.273798\\
 \hline
26--30 & 0.249741 &  0.267864 & 0.204698 & 0.292317 & 0.209712\\
 \hline
31--35 & 0.278560 & 0.220264 & 0.282694& 0.221687 & 0.268559\\
  \hline
36--38 & 0.233460 & 0.270419 & 0.223050\\
 \hline
\end{tabular}
\caption{Structure of an Advanced LIGO-like coating optimized jointly for dichroic 
  operation and thermal noise.  Thickness of each layer is given in units of wavelength 
  (for light with vacuum-wavelength of 1064\,nm) are listed here for the 38 layers. Note that $l_{1,3,5,\ldots}$ are $\mathrm{SiO}_2$ 
  layers, while $l_{2,4,6,\ldots}$ are $\mathrm{Ta}_2\mathrm{O}_5$ layers.
 \label{tab:asv}}
\end{table}

In Table~\ref{tab:asv},  we provide the structure of the coating optimized jointly for dichroic operation 
and thermal noise (baseline parameter, neglecting penetration).

\end{document}